\documentstyle[preprint,aps,epsf,psfig,eqsecnum]{revtex}
\begin{document}
\def\be{\begin{eqnarray}}
\def\en{\end{eqnarray}}
\def\non{\nonumber}
\def\la{\langle}
\def\ra{\rangle}
\def\ep{\varepsilon}
\def\hep{\hat{\varepsilon}}
\def\ek{{\vec{\ep}_\perp\cdot\vec{k}_\perp}}
\def\epp{{\vec{\ep}_\perp\cdot\vec{P}_\perp}}
\def\kp{{\vec{k}_\perp\cdot\vec{P}_\perp}}
\def\lsim{ {\ \lower-1.2pt\vbox{\hbox{\rlap{$<$}\lower5pt\vbox{\hbox{$\sim$}
}}}\ } }
\def\gsim{ {\ \lower-1.2pt\vbox{\hbox{\rlap{$>$}\lower5pt\vbox{\hbox{$\sim$}
}}}\ } }
\def\dk{\partial\!\cdot\!K}
\def\pr{{\sl Phys. Rev.}~}
\def\prl{{\sl Phys. Rev. Lett.}~}
\def\pl{{\sl Phys. Lett.}~}
\def\np{{\sl Nucl. Phys.}~}
\def\zp{{\sl Z. Phys.}~}

\draft
\title{
\begin{flushright}
{\normalsize ~~~IP-ASTP-02-97
}
\end{flushright}
\Large\bf
A Phenomenological Analysis of Heavy Hadron Lifetimes
}
\author{{\bf Hai-Yang Cheng}
}
\vskip 2.0cm
\address{Institute of Physics, Academia Sinica, Taipei, Taiwan 115, 
Republic of China}
%
\vskip 1.0 cm
\date{June, 1997}
\maketitle
\begin{abstract}
 A phenomenological analysis of lifetimes of bottom and charmed hadrons
within the framework of the heavy quark expansion is performed. The baryon 
matrix element is evaluated using the bag model and the nonrelativistic 
quark model.  We find that bottom-baryon lifetimes follow the pattern
$\tau(\Omega_b)\simeq\tau(\Xi_b^-)>\tau(\Lambda_b)\simeq\tau(\Xi_b^0)$.
However, neither the lifetime ratio $\tau(\Lambda_b)/\tau( B_d)$ nor the
absolute decay rates of the $\Lambda_b$ baryon and $B$ mesons can be
explained. One way of solving both difficulties is to allow the presence 
of linear $1/m_Q$ corrections by scaling the inclusive nonleptonic width
with the fifth power of the hadron mass $m_{H_Q}$ rather than the heavy 
quark mass $m_Q$. The hierarchy of bottom baryon lifetimes is dramatically
modified to $\tau(\Lambda_b)>\tau(\Xi_b^-)>\tau(\Xi_b^0)>\tau(
\Omega_b)$: The longest-lived $\Omega_b$ among bottom baryons in the 
OPE prescription now becomes shortest-lived. The replacement of 
$m_Q$ by $m_{H_Q}$ in nonleptonic widths is natural and justified
in the PQCD-based factorization approach formulated in terms of hadron-level
kinematics. For inclusive charmed baryon decays, we 
argue that since the heavy quark expansion does not converge, local 
duality cannot be tested in this case. We show that while the ansatz of 
substituting the heavy quark mass by the hadron mass provides a much better
description of the charmed-baryon lifetime {\it ratios}, it appears unnatural
and unpredictive for describing the {\it absolute} inclusive decay rates
of charmed baryons, contrary to the bottom case.

\vskip 1.0 true cm
PACS numbers: 12.39.Hg, 12.39.Jh, 13.20.He, 13.30-a
\end{abstract}
\newpage
\baselineskip .29in

\section{Introduction}
   The lifetime differences among the charmed mesons $D^+,~D^0$ and charmed
baryons have been studied extensively both experimentally and theoretically
since late seventies. It was realized very early that the naive parton
model gives the same lifetimes for all heavy particles containing a heavy
quark $Q$ and that the underlying mechanism for the decay width
differences and the lifetime hierarchy of heavy hadrons comes mainly from
the nonspectator effects like $W$-exchange and Pauli interference
due to the identical quarks produced in heavy quark decay and in the
wavefunction (for a review, see \cite{Bigi,BS94}). The nonspectator 
effects were expressed in eighties
in terms of local four-quark operators by relating the total widths to
the imaginary part of certain forward scattering amplitudes 
\cite{Bilic,Guberina,SV}. (The nonspectator effects for charmed baryons were
first studied in \cite{Ruckl}.)
With the advent of heavy quark effective theory (HQET), it was recognized
in early nineties that nonperturbative corrections to the parton picture
can be systematically expanded in powers of $1/m_Q$ \cite{Bigi92,BS93}.
Subsequently, it was demonstrated that this $1/m_Q$ expansion is applicable 
not only to global quantities such as lifetimes, but also to local 
quantities, e.g. the lepton spectrum in the semileptonic decays of heavy
hadrons \cite{Bigi93}.
Therefore, the above-mentioned phenomenological work in eighties 
acquired a firm theoretical footing in nineties, namely the heavy
quark expansion (HQE), which is a generalization of the operator product
expansion (OPE) in $1/m_Q$. Within this QCD-based framework, some 
phenomenological assumptions can be turned into some coherent and 
quantitative statements and nonperturbative
effects can be systematically studied. As an example, consider the baryon
matrix element of the two-quark operator $\la\Lambda_b|\bar bb|\Lambda_b\ra$.
The conventional quark-model evaluation of this matrix element is 
model-dependent:
\be
{\la\Lambda_b|\bar bb|\Lambda_b\ra\over 2m_{\Lambda_b}}=\cases{ 1 & NQM;   \cr
\int d^3r\,[u_b^2(r)-v_b^2(r)] & bag~model,   \cr}
\en
where $u(r)$ and $v(r)$ are the large and small components, respectively, 
of the quark wavefunction. However, the matrix element (1.1), which is
equal to unity in the nonrelativistic quark model (NQM), becomes smaller
in the bag model due to the contribution from the lower component of the
quark wavefunction. In the HQE approach, it is given by [see Eq.~(2.8)
below]
\be
{\la\Lambda_b|\bar bb|\Lambda_b\ra\over 2m_{\Lambda_b}}=\,1+{1\over 2m_b^2}
\left({\la\Lambda_b|\bar b(iD_\perp)^2b|\Lambda_b\ra\over 2m_{\Lambda_b}}
\right)+{1\over 4m_b^2}\left({\la\Lambda_b|\bar b\sigma\cdot Gb|\Lambda_b
\ra\over 2m_{\Lambda_b}}\right)+{\cal O}(1/m_b^3),
\en
with $D_\perp^\mu=\partial^\mu-v^\mu v\!\!\cdot\!\! D$.
This expression is not only model independent but also contains 
nonperturbative kinetic and chromomagnetic effects which are either
absent or overlooked in the earlier quark-model calculations.

   Based on the OPE approach for the analysis of inclusive weak decays, 
predictions for the ratios of bottom hadron lifetimes have been made 
by several groups. The first correction to bottom hadron lifetimes is 
of order $1/m_b^2$ and it is model independent \cite{Neubert97}:
\be
{\tau(B^-)\over\tau(B_d)} &=& 1+{\cal O}(1/m_b^3),   \non \\
{\tau(B_s)\over\tau(B_d)} &=& (1.00\pm 0.01)+{\cal O}(1/m_b^3),   \non \\
{\tau(\Lambda_b)\over\tau(B_d)} &=& 0.98+{\cal O}(1/m_b^3).
\en
The $1/m_b^2$ corrections are small and essentially canceled
out in the lifetime ratios.
Nonspectator effects in inclusive decays due to the Pauli interference and
$W$-exchange contributions account for $1/m_b^3$ corrections and they
have two eminent features: First, the estimate of nonspectator effects is
model dependent; the hadronic four-quark matrix elements are usually 
evaluated by assuming the factorization approximation for mesons and 
the quark model for baryons. Second, $1/m_b^3$ corrections can be quite
significant due to a phase-space enhancement by a factor of $16\pi^2$. 
Predictions made in \cite{Bigi94} for lifetime ratios of bottom hadrons are
\be
{\tau(B^-)\over\tau(B_d)}=\,1.0+0.05\left({f_B\over 200\,{\rm MeV}}\right)
^2,~~~~~~~{\tau(\Lambda_b)\over \tau(B_d)}\gtrsim 0.9\,.
\en
Experimentally \cite{LEP}, while the $B^-$ and $B_d$ lifetimes are very 
close, it appears that the $\Lambda_b$ lifetime is significantly shorter 
than the $B$ meson one:
\be
{\tau(B^-)\over\tau(B_d)}=\,1.06\pm 0.04\,,~~~~~
{\tau(\Lambda_b)\over\tau(B_d)}=\,0.79\pm 0.06~~~~{\rm (world~average)}.
\en
It should be mentioned that while the world average value for 
$\tau(\Lambda_b)/\tau(B_d)$ is dominated by LEP experiments \cite{LEP}, 
the CDF experiment alone yields \cite{Tseng}
\be
{\tau(\Lambda_b)\over\tau(B_d)}=\,0.87\pm 0.11~~~~{\rm (CDF)}.
\en
It is thus important to fully settle down the experimental 
situation in the near future.
Evidently, the conflict between experiment (1.5) and theoretical 
expectations from (1.3) or (1.4) is striking and intriguing. This has 
motivated several
subsequent studies trying to understand the enhancement of the $\Lambda_b$ 
decay rate \cite{Neubert97,Uraltsev,Rosner,Altarelli,Col}.
For example, a model-independent analysis in \cite{Neubert97} gives
\be
{\tau(B^-)\over\tau(B_d)} &\simeq& 1+0.03B_1+0.004B_2-0.70\ep_1+0.20\ep_2, 
\non \\
{\tau(\Lambda_b)\over\tau(B_d)} &\simeq &0.98-0.17\ep_1+0.20\ep_2-(0.012+
0.021\tilde B)r,   
\en
where $\ep_i,~B_i,~\tilde B,~r$ are the hadronic parameters to be introduced
below in Sec.~II. Note that while the ratio $\tau(B^-)/\tau(B_d)$ is predicted
to be greater than unity in \cite{Bigi94} [see (1.4)], it was argued in
\cite{Neubert97} that the unknown nonfactorizable contributions in (1.7)
characterized by $\ep_i$ make it impossible to have reliable predictions
on the magnitude of the lifetime ratio and even the sign of corrections.
Since the measured ratio of $\tau(B^-)/\tau(B_d)$ 
is very close to unity, it follows from (1.7) that $\ep_1\approx 0.3\,\ep_2$ 
\cite{Neubert97}. Then it is clear that the 
data for the ratio $\tau(\Lambda_b)/\tau(B_d)$ cannot be accommodated 
by the theoretical prediction (1.7) without invoking a too large value of 
$r$ or $\tilde B$, which is expected to be 
order unity. It is reasonable to conclude that the $1/m_b^3$ corrections 
in the heavy quark expansion do not suffice to describe the observed lifetime 
differences between $\Lambda_b$ and $B_d$.

In order to employ the OPE approach to compute inclusive weak decays of heavy
hadrons, some sort of quark-hadron duality has to be assumed (for an extensive
discussion of quark-hadron duality and its violation, see \cite{Falk,BM}). 
Consider the
inclusive semileptonic decay. The OPE cannot be carried out on the physical
cut in the complex $v\!\cdot\!q$ plane since $T^{\mu\nu}$, the 
time-ordered product of two currents, along the physical cut is dominated 
by physical intermediate hadron states which are nonperturbative in nature.
To compute $T^{\mu\nu}$ or the Wilson coefficients by perturbative
QCD, the OPE has to be performed in the unphysical region far away from the
physical cut. The question is then how to relate the operator product 
expansion for $T^{\mu\nu}$ in the unphysical region to the physical 
quantities in the physical Minkowski space. Since the physically 
observable quantity is related 
to the imaginary part of $T^{\mu\nu}$, it can be reliably computed by 
deforming the contour of integration into the unphysical region
\cite{Chay,Falk}, provided that the physical quantity involves certain
integrals of $T^{\mu\nu}$ in the physical region.
This procedure is called ``global duality"
\cite{Falk}. Global quark-hadron duality also means that the hadronic
cross section is dual or matching to the OPE-based quark cross section.
However, unlike the total cross section
in $e^+e^-$ annihilation, there is a small portion of the contour near the
physical cut where global duality can no longer be applied. As stressed in
\cite{Falk}, one must resort to local duality to justify the use of the OPE
in this small region. Fortunately, the contribution is of order $\Lambda_{
\rm QCD}/m_Q$ and can be neglected for quantities smeared over an energy scale
of order $\Lambda_{\rm QCD}$.

    Global quark-hadron duality for inclusive semileptonic decays, namely
the matching between the hadronic and OPE-based expressions for decay widths
or smeared spectra in semileptonic $B$ and $\Lambda_b$ decays has been 
explicitly proved to the first two terms in $1/m_b$ expansion and the 
first order in $\alpha_s$ in the Shifman-Voloshin (SV) limit
\cite{Boyd}. The hadronic decay rate is calculated by summing over all 
allowed exclusive decay channels. In the SV limit for $B$ meson decays via
$b\to c$ transitions, the dominant hadronic final states are the $D$ 
and $D^*$. (At zero recoil, the quark-mixing-favored semileptonic decays
of a $B$ meson in the heavy quark limit can only produce a $D$ or 
$D^*$ meson \cite{Boyd,Yan}.) The exclusive decay rates or distributions
for $B\to (D+D^*)\ell\bar\nu$ depend on hadron masses, whereas the inclusive
decay rates evaluated by the OPE depend on quark masses. Global duality
is then proved by showing explicitly the equality of inclusive
and exclusive decay rates. Note that this proof of global
duality in QCD is valid only in the SV limit. Beyond this limit, it becomes
difficult to sum over all allowed exclusive semileptonic decay channels and
evaluate all of them. It was shown
recently in \cite{Yu} that a proof of quark-hadron global 
duality in the general kinematic region to order $(\Lambda_{\rm QCD}/m_B)^2$
can be achieved in the PQCD-based factorization approach, which is
formulated in terms of meson-level kinematics rather than the quark-level
one. It was demonstrated explicitly in \cite{Yu} that the integrated 
quark-level spectrum equals to the hadron-level spectrum and that linear 
$1/m_b$ corrections to the total decay rate are nontrivially canceled out, in
agreement with the OPE expectation \cite{Bigi92,BS93}. \footnote{The absence
of linear $1/m_b$ corrections to decay widths is trivial in the SV limit
since the inclusive decay rates depend on $\Delta M=m_B-m_D$ rather
than $m_B$, and $\Delta M=\delta m+{\cal O}(1/m_b^2)$ with $\delta 
m=m_b-m_c$ \cite{Boyd}.}

   Unlike the semileptonic inclusive decays in which the use of the OPE is
validated by deforming the contour away from the physical cut, 
it is pointed out in \cite{Falk} that there is no external momentum $q$ 
in inclusive nonleptonic decays which allows analytic continuation into
the complex plane. Therefore, the OPE is {\it a priori} not justified in this 
case and local duality has to be invoked in order to apply
the OPE directly in the physical region. It is obvious that local 
quark-hadron duality is less firm and secure than global duality, although 
its validity has been proved to the first two terms in $1/m_Q$ expansion
and the first order in $\alpha_s$ in the SV limit under the factorization 
hypothesis \cite{Deandrea}. It should be stressed that quark-hadron 
duality is {\it exact} in the heavy quark limit, but its systematical 
$1/m_Q$ expansion is still lacking. It is very likely that $1/m_Q$
corrections to quark-hadron duality behave differently for inclusive
semileptonic and nonleptonic decays.
Motivated by the conflict between theory and experiment for the lifetime
ratio $\tau(\Lambda_b)/\tau(B_d)$, it was suggested in \cite{Altarelli} that
the assumption of local duality is not correct for nonleptonic inclusive
width and that the presence of linear $1/m_b$ corrections is strongly 
indicated by the data. Moreover, the $1/m_b$ corrections are well described
by the simple ansatz that the heavy quark mass $m_Q$ is replaced by the
decaying hadron mass in the $m_Q^5$ factor in front of all nonleptonic
widths. It is easily seen that the factor $(m_B/m_{\Lambda_b})^5=0.73$ is
very close to the observed value of $\tau(\Lambda_b)/\tau(B_d)$ .
Under this ansatz, a much better description of lifetimes of bottom
and charmed hadrons was shown in \cite{Altarelli}. Irrespective of the
lifetime ratio problem, there is another important reason why this ansatz
is welcome. The absolute decay rate of the $B$ meson predicted in the 
OPE approach is at least 20\% smaller than the experimental value (see 
Sec.~III below). We shall
show in Sec.~III that the discrepancy between theory and experiment is
greatly improved when the nonleptonic width scales with $m_B^5$. 
    
   In the aforementioned factorization approach of \cite{Yu}, the 
nonleptonic width $\Gamma_{\rm NL}^{\rm had}$ of bottom hadrons
scales with $m_{H_b}^5$. Local duality means that a replacement of meson-level
kinematics by quark kinematics, for example, $m_{H_b}=m_b(1+\bar\Lambda_{H_b}/
m_b+\cdots),\cdots$ etc., will turn $\Gamma_{\rm NL}^{\rm had}$ into 
$\Gamma_{\rm NL}^{\rm OPE}$, the OPE-based decay rate. 
Consequently, the relation 
between violation of local duality and the above-mentioned ansatz will
become natural in the factorization approach.

    In the present paper we will study nonspectator effects in inclusive
nonleptonic and semileptonic decays and 
analyze the lifetime pattern of heavy hadrons. In 
particular, we focus on the lifetimes of heavy baryons and study
the implications of broken local duality. We will demonstrate that the
lifetime hierarchy of bottom baryons is dramatically modified when the
quark mass is replaced by the hadron mass in nonleptonic widths. 
The layout of this paper is organized as follows. In Sec.~II we give 
general heavy quark expansion expressions for inclusive nonleptonic and
semileptonic widths and pay attention
to the evaluation of baryon four-quark matrix elements and the nonperturbative 
parameter $\lambda_2$ for baryons. We then study bottom-hadron lifetimes 
in Sec.~III and apply the ansatz mentioned above. In Sec.~IV we examine 
the applicability of the same prescription to charmed baryon decays. 
Discussions and conclusions are given in Sec.~V.

\section{Framework}
   In this section we write down the general expressions for the 
inclusive decay widths of heavy hadrons and evaluate the relevant hadronic 
matrix elements. It is known that the inclusive decay rate is governed by 
the imaginary part of an effective nonlocal forward transition operator $T$. 
When the energy released in the decay is large enough, the nonlocal effective
action can be recast as an infinite series of local operators with
coefficients containing inverse powers of the heavy quark mass $m_Q$.
Under this heavy quark expansion, the inclusive
nonleptonic decay rate of a heavy hadron $H_Q$ containing a heavy quark $Q$
is given by \cite{Bigi92,BS93}
\be
\Gamma_{\rm NL}(H_Q) &=& {G_F^2m_Q^5\over 192\pi^3}N_c\,\xi\,
{1\over 2m_{H_Q}}
\Bigg\{ \left(c_1^2+c_2^2+{2c_1c_2\over N_c}\right)\times \non \\
&& \Big[ \big(I_0(x,0,0)+I_0
(x,x,0)\big)\la H_Q|\bar QQ|H_Q\ra   \non \\
&& -{1\over m_Q^2}\big(I_1(x,0,0)+I_1(x,x,0)\big) \la H_Q|\bar Q\sigma\cdot G 
Q|H_Q\ra \Big]   \non \\
&& -{4\over m_Q^2}\,{2c_1c_2\over N_c}\,\big(I_2(x,0,0)+I_2(x,x,0)\big)
\la H_Q|\bar Q\sigma\cdot G Q|H_Q\ra\Bigg\}   \non \\
&& +{1\over 2m_{H_Q}}\la H_Q|{\cal L}_{\rm nspec}|H_Q\ra+{\cal O}(1/m_Q^4),
\en
where $\sigma\cdot G=\sigma_{\mu\nu}G^{\mu\nu}$, $c_1,~c_2$ are Wilson
coefficient functions, $N_c=3$ is the number of color, the factor $\xi$ 
takes care of the relevant CKM matrix elements,
for example, $\xi=|V_{cb}V_{ud}|^2$ for quark-mixing-favored bottom decay,
$I_0,~I_1$ and $I_2$ are phase-space factors:
\be
I_0(x,0,0) &=& (1-x^2)(1-8x+x^2)-12x^2\ln x,   \non \\
I_1(x,0,0) &=& {1\over 2}(2-x{d\over dx})I_0(x,0,0)=(1-x)^4,   \non \\
I_2(x,0,0) &=& (1-x)^3
\en
for $b\to c\bar ud$ ($x=m_c^2/m_b^2$) or $c\to su\bar d$ ($x=m_s^2/m_c^2$)
transition and
\be
I_0(x,x,0) &=& v(1-14x-2x^2-12x^3)+24x^2(1-x^2)\ln {1+v\over 1-v},   \non \\
I_1(x,x,0) &=& {1\over 2}(2-x{d\over dx})I_0(x,x,0),   \non \\
I_2(x,x,0) &=& v(1+{x\over 2}+3x^2)-3x(1-2x^2)\ln {1+v\over 1-v}
\en
for $b\to cc\bar s$ transition with $v\equiv\sqrt{1-4x}$.

  The dimension-six four-quark operators ${\cal L}_{\rm nspec}$ in (2.1) 
describe nonspectator effects
in inclusive decays of heavy hadrons and are given by 
\cite{Bilic,Guberina,SV}
\be
{\cal L}_{\rm nspec} &=& {G^2_Fm_Q^2\over 2\pi}\,\xi\,(1-x)^2\Big\{
(c_1^2+c_2^2)(\bar Q Q)(\bar q_1q_1)+2c_1c_2(\bar Qq_1)(\bar q_1 Q)\Big\}
\non \\
&-& {G_F^2m_Q^2\over 6\pi}\,\xi\Bigg\{ c_1^2(1-x)^2\big[ (1+
{x\over 2})(\bar QQ)(\bar q_2q_2)-(1+2x)\bar Q^\alpha(1-\gamma_5)q_2^\beta
\bar q_2^\beta(1+\gamma_5)Q^\alpha\big]   \non \\
&+& (2c_1c_2+N_cc_2^2)(1-x)^2\big[ (1+{x\over 2})(\bar Qq_2)(\bar q_2Q)-
(1+2x)\bar Q(1-\gamma_5)q_2\bar q_2(1+\gamma_5)Q\big]\Bigg\}   \non \\
&-& {G_F^2m_Q^2\over 6\pi}\,\xi\Bigg\{
c_1^2\sqrt{1-4x}\,\big[ (1-x)(\bar Q Q)(\bar q_3q_3)-(1+2x)\bar Q^\alpha(1-
\gamma_5)q_3^\beta\bar q_3^\beta (1+\gamma_5)Q^\alpha]   \non \\
&+& (2c_1c_2+N_cc_2^2)\sqrt{1-4x}\,\big[ (1-x)(\bar Qq_3)(\bar q_3Q)-
(1+2x)\bar Q(1-\gamma_5)q_3\bar q_3(1+\gamma_5)Q\big]\Bigg\}   \non \\
&-& {G_F^2m_Q^2\over 6\pi}\,\xi\Bigg\{
c_2^2(1-x)^2\big[ (1+
{x\over 2})(\bar QQ)(\bar q_3q_3)-(1+2x)\bar Q^\alpha(1-\gamma_5)q_3^\beta
\bar q_3^\beta(1+\gamma_5)Q^\alpha\big]   \non \\
&+& (2c_1c_2+N_cc_1^2)(1-x)^2\big[ (1+{x\over 2})(\bar Qq_3)(\bar q_3Q)-
(1+2x)\bar Q(1-\gamma_5)q_3\bar q_3(1+\gamma_5)Q\big] \Bigg\},
\en
where $(\bar q'q)\equiv \bar q'\gamma_\mu(1-\gamma_5)q$, and $\alpha,~\beta$
are color indices.
Note that for charm decay, $Q=c,~q_1=u,~q_2=d$ and $q_3=s$ and for bottom
decay, $Q=b,~q_1=u,~q_2=d,~q_3=s$. The last term in (2.4) is due to
the constructive interference of the $s$ quark and hence it occurs only 
in charmed baryon decays. The 
third term in (2.4) exists only in bottom decays with $c\bar c$ intermediate 
states. For inclusive nonleptonic decays of heavy mesons, the first term 
in (2.4) corresponds to a Pauli interference and the second and third 
terms to $W$-exchange contributions. For heavy baryon decays, the first term 
is a $W$-exchange contribution and the rest are interference terms. The 
phase-space suppression factors
e.g. $(1-x)^2,~\sqrt{1-4x},\cdots$ etc. in (2.4) are derived in 
\cite{Chernyak,Neubert97}. 

   Several remarks are in order. (i) There is no linear $1/m_Q$ corrections 
to the inclusive decay rate due to the lack of gauge-invariant dimension-four
operators \cite{Chay,Bigi92}, a consequence known as Luke's 
theorem \cite{Luke}. Nonperturbative corrections start at order $1/m_Q^2$.
(ii) It is clear from Eqs.~(2.1) and (2.4) that there is a two-body 
phase-space enhancement factor of $16\pi^2$ for nonspectator effects
relative to the three-body phase space for heavy quark decay. This implies 
that nonspectator effects, being of order $1/m_Q^3$, are comparable to and 
even exceed the $1/m_Q^2$ terms. (iii) For charmed 
meson decay, the $1/N_c$ correction to $\Gamma_{\rm NL}$ characterized by 
the term $(2c_1c_2/N_c)\la H_c|\bar cc|H_c\ra$ is found to be compensated
by the nonperturbative gluonic effect [i.e. the term proportional to 
$I_2(x,0,0)$]. This cancellation is small for $B$ meson decay due to the
smallness of $1/m_b^2$. This indicates that the rule of discarding $1/N_c$ 
terms \cite{Buras86}
is operative in charm decays but not so for the $B$ meson case. (iv) Thus
far the Wilson coefficients and four-quark operators in Eq.~(2.4) are
renormalized at the heavy quark mass scale. Sometimes the so-called hybrid
renormalization \cite{SV,SV87} is performed to evolve the four-quark operators 
(not the Wilson coefficients !) from $m_Q$ down to a low energy scale, say, a 
typical hadronic scale $\mu_{\rm had}$. The underlying reason is that the
factorizable approximation for meson matrix elements and the quark model
for baryon matrix elements are believed to be more reliable at the scale 
$\mu_{\rm had}$. The evolution from $m_Q$ down to $\mu_{\rm had}$ will in
general introduce new structures such as penguin operators. However, in 
the present paper we will follow \cite{Neubert97} to employ (2.1) and 
(2.4) as our starting point for describing inclusive weak decays
since it is equivalent to first evaluating the four-quark
matrix elements renormalized at the $m_Q$ scale and then relating them
to the hadronic matrix elements renormalized at $\mu_{\rm had}$ through
the renormalization group equation, provided that the effect of penguin 
operators is neglected.

   For inclusive semileptonic decays, apart from the heavy quark decay
contribution there is an additional nonspectator effect in charmed-baryon
semileptonic decay originating from the Pauli interference of the $s$
quark \cite{Voloshin}.  It is now ready to deduce
the inclusive semileptonic widths from (2.1) 
and the last term in (2.4) by putting $c_1=1$, $c_2=0$ and $N_c=1$:
\be
\Gamma_{\rm SL}(H_Q) &=& {G_F^2m_Q^5\over 192\pi^3}|V_{\rm CKM}|^2\,{
\eta(x,x_\ell,0)\over 
2m_{H_Q}}\Big[ I_0(x,0,0)\la H_Q|\bar QQ|H_Q\ra-{1\over m_Q^2}\,I_1(x,0,0)
\la H_Q|\bar Q\sigma\cdot G Q|H_Q\ra \Big]   \non \\
&-& {G_F^2m_c^2\over 6\pi}|V_{cs}|^2{1\over 2m_{H_c}}(1-x)^2\Big[(1+{x\over 2})
(\bar c s)(\bar s c)-(1+2x)\bar c(1-\gamma_5)s\bar s(1+\gamma_5)c\Big],
\en
where $\eta(x,x_\ell,0)$ with $x_\ell=(m_\ell/m_Q)^2$ is the QCD radiative 
correction to the semileptonic decay rate. Its general analytic
expression is given in \cite{Hokim}. The special case $\eta(x,0,0)$ is given
in \cite{Nir} and it can be approximated numerically by \cite{Kim}:
\be
\eta(x,0,0)\cong\,1-{2\alpha_s\over 3\pi}\left[ (\pi^2-{31\over 4})(1-
\sqrt{x})^2+{3\over 2}\right].
\en
With $x=0$ and the replacement $\alpha_s\to {3\over 4}\alpha$, (2.6) is
reduced to the well-known QED correction to the muon decay. The second term
in Eq.~(2.5) occurs only in the semileptonic decay of $\Xi_c$ and $\Omega_c$
baryons.

    We next turn to the 2-body matrix elements $\la H_Q|\bar QQ|H_Q\ra$. 
The use of the equation of motion
\be
\bar QQ=\bar Qv\!\!\!/Q+{1\over 2m_Q^2}\bar Q(i{D_\perp})^2Q+{1\over 4m_Q^2}
\bar Q\sigma\cdot GQ+{\cal O}(1/m_Q^3),
\en
with $D_\perp^\mu=\partial^\mu-v^\mu v\!\cdot\!D$, leads to
\be
{\la H_Q|\bar QQ|H_Q\ra\over 2m_{H_Q}}=1-{K_H\over 2m_Q^2}+{G_H\over 2m_Q^2},
\en
with
\be
&& K_H\equiv -{1\over 2m_{H_Q}}\la H_Q|\bar Q(i{D_\perp})^2Q|H_Q\ra=
-\lambda_1,   \non \\
&& G_H\equiv{1\over 2m_{H_Q}}\la H_Q|\bar Q{1\over 2}\sigma\cdot 
GQ|H_Q\ra=d_H\lambda_2.   
\en
The mass of the heavy hadron $H_Q$ is then of the form
\be
m_{H_Q}=\,m_Q+\bar\Lambda_{H_Q}-{\lambda_1\over 2m_Q}-{d_H\lambda_2\over 2m_Q},
\en
where the three nonperturbative HQET parameters $\bar\Lambda_{H_Q},~
\lambda_1$, $\lambda_2$ are independent of the heavy quark mass and in general 
$\bar\Lambda_{H_Q}$ is different for different heavy hadrons.
Since $\sigma\cdot G\sim \vec{S}_Q\cdot 
\vec{B}$ and since the chromomagnetic field is produced by the light cloud
inside the heavy hadron, it is clear that $\sigma\cdot G$ is proportional to
$\vec{S}_Q\cdot\vec{S}_\ell$, where $\vec{S}_Q~(\vec{S}_\ell)$ is the spin 
operator of the heavy quark (light cloud). More precisely,
\be
d_H &=& -\la H_Q|4\vec{S}_Q\cdot\vec{S}_\ell|H_Q\ra   \non \\
&=& -2[S_{\rm tot}(S_{\rm tot}+1)-S_Q(S_Q+1)-S_\ell(S_\ell+1)].
\en
Therefore, $d_H=3$ for $B,~D$ mesons, $d_H=-1$ for $B^*,~D^*$ mesons, $d_H=0$
for the antitriplet baryon $T_Q$, $d_H=4$ for the spin-${1\over 2}$ 
sextet baryon $S_Q$ and $d_H=-2$ for the spin-${3\over 2}$ sextet baryon 
$S^*_Q$. It follows from (2.10) that
\be
\lambda_2^{\rm meson} &=& {1\over 4}(m^2_{P^*}-m_P^2)=\cases{ 0.12\,{\rm 
GeV}^2& for~B~meson;   \cr  0.14\,{\rm GeV}^2 & for~D~meson,   \cr}   \non \\
\lambda_2^{\rm baryon} &=& {1\over 6}(m^2_{S^*_Q}-m^2_{S_Q}).
\en
The values of $\lambda_2^{\rm baryon}$ will be fixed later.
As for the kinetic energy parameter $\lambda_1$ we use
\cite{Neubert96}
\be
\lambda_1^{\rm meson}\sim \lambda_1^{\rm baryon}=-(0.4\pm 0.2)\,{\rm GeV}^2.
\en
This leads to
\be
m_b-m_c &=& (\la m_B\ra-\la m_D\ra)\left(1-{\lambda_1\over 2\la m_B\ra\la 
m_D\ra}\right)   \non \\
&=& (3.40\pm 0.03)\,{\rm GeV},   
\en
where $\la m_P\ra={1\over 4}(m_P+3m_{P^*})$ denotes the spin-averaged meson
mass.

   We will follow \cite{Neubert97} to parametrize the hadronic matrix 
elements in a model-independent way.
    For meson matrix elements of four-quark operators, we follow 
\cite{Neubert97} to define the parameters $B_i$ and $\ep_i$:
\be
&& \la \bar B_q|(\bar bq)(\bar qb)|\bar B_q\ra = \,f_{B_q}^2m_{B_q}^2 B_1, 
\non \\
&& \la \bar B_q|\bar b(1-\gamma_5)q\bar q(1+\gamma_5)b|\bar B_q\ra =\,f_{
B_q}^2m_{B_q}^2 B_2, \non \\
&& \la \bar B_q|(\bar b\,t^aq)(\bar q\,t^ab)|
\bar B_q\ra = \,f_{B_q}^2m_{B_q}^2 \ep_1, \non \\
&& \la \bar B_q|\bar b\,t^a(1-\gamma_5)q\bar q\,t^a
(1+\gamma_5)b|\bar B_q\ra = \,f_{B_q}^2m_{B_q}^2 \ep_2,
\en
where $(\bar q't^aq)\equiv\bar q't^a\gamma_\mu(1-\gamma_5)q$ and 
$t^a=\lambda^a/2$. Under the factorization approximation, 
$B_i$ and $\ep_i$ are given by $B_i=1$ and 
$\ep_i=0$, but they will be treated as free parameters here. As a 
consequence of (2.15), we obtain
\be
\la \bar B_q|(\bar bb)(\bar qq)|\bar B_q\ra &=& f_{B_q}^2m_{B_q}^2\left({1
\over 3}B_1+2\ep_1\right),   \non \\
\la \bar B_q|\bar b^\alpha(1-\gamma_5)q^\beta\bar q^\beta(1+\gamma_5)b^\alpha
|\bar B_q\ra &=& f_{B_q}^2m_{B_q}^2\left({1\over 3}B_2+2\ep_2\right).   
\en

    As for the baryon matrix elements of four-quark operators we have to rely
on the quark model. We first consider the MIT bag model \cite{MIT} and 
define three four-quark overlap integrals:
\be
a_q &=& \int d^3r[\,u_q^2(r)u_Q^2(r)+v_q^2(r)v_Q^2(r)],   \non \\
b_q &=& \int d^3r[\,u_q^2(r)v_Q^2(r)+v_q^2(r)u_Q^2(r)],   \non \\
c_q &=& \int d^3r\,u_q(r)v_q(r)u_Q(r)v_Q(r),
\en
which are expressed in terms of the large and small components $u(r)$ and 
$v(r)$, respectively, of the quark wavefunction. For the antitriplet heavy 
baryon $T_Q$ or the sextet heavy baryon $\Omega_Q$ (recall that only 
the $\Omega_c^0$ and
$\Omega_b^-$ of the sextet baryons decay weakly), the four baryon matrix 
elements
\be
&& \la T_Q|(\bar Qq)(\bar qQ)|T_Q\ra,~~~~~~\la T_Q|(\bar QQ)(\bar qq)|T_Q\ra,
\non \\   && \la T_Q|\bar Q(1-\gamma_5)q\bar q(1+\gamma_5)Q|T_Q\ra,~~~
\la T_Q|\bar Q^\alpha(1-\gamma_5)q^\beta\bar q^\beta(1+\gamma_5)Q^\alpha|
T_Q\ra    \non 
\en
are not all independent. First of all, we have
\be
\la T_Q|(\bar Qq)(\bar qQ)|T_Q\ra &=& -(a_q+b_q)(2m_{T_Q}),   \non \\
\la \Omega_Q|(\bar Qs)(\bar sQ)|\Omega_Q\ra &=& -{1\over 3}(18a_s+2b_s+
32c_s)(2m_{\Omega_Q}),
\en
(see e.g., Ref.~\cite{CT} for the technical detail of the bag model
evaluation), where we have taken into account the fact that there are two
valence $s$ quarks in the wavefunction of the $\Omega_Q$.
Second, since the color wavefunction for a baryon is totally antisymmetric,
the matrix element of $(\bar QQ)(\bar qq)$ is the same as that of $(\bar Qq)
(\bar qQ)$ except for a sign difference. Thus we follow \cite{Neubert97} to
define a parameter $\tilde B$
\be
\la T_Q|(\bar QQ)(\bar qq)|T_Q\ra  &=& -\tilde{B}\la T_Q|(\bar Q q)(\bar qQ)
|T_Q\ra,    \non \\
\la \Omega_Q|(\bar QQ)(\bar ss)|\Omega_Q\ra  &=& -\tilde{B}\la \Omega_Q|(
\bar Q s)(\bar sQ)|\Omega_Q\ra,
\en
so that $\tilde B=1$ in the valence-quark approximation. Third, it is 
straightforward to show that
\be
\la T_Q|\bar Q^\alpha\gamma_\mu\gamma_5 Q^\beta\bar q^\beta\gamma^\mu(1-
\gamma_5)q^\alpha|T_Q\ra &=& 0,   \non  \\
\la \Omega_Q|\bar Q^\alpha\gamma_\mu\gamma_5 Q^\beta\bar q^\beta\gamma^\mu(1-
\gamma_5)q^\alpha|\Omega_Q\ra &=& 4\left(a-{b\over 3}\right)(2m_{\Omega_Q}).
\en
The first relation in (2.20) is actually a model-independent consequence of 
heavy-quark spin symmetry \cite{Neubert97}. Since
\be
\bar Q^\alpha\gamma_\mu\gamma_5 Q^\beta\bar q^\beta\gamma^\mu(1-\gamma_5)
q^\alpha=-\bar Q(1-\gamma_5)q\bar q(1+\gamma_5)Q-{1\over 2}(\bar Qq)(\bar 
q Q),
\en
it follows from (2.20) that
\be
\la T_Q|\bar Q^\alpha(1-\gamma_5)q^\beta\bar q^\beta(1+\gamma_5)Q^\alpha|
T_Q\ra  &=& -\tilde B\la T_Q|\bar Q(1-\gamma_5)q\bar q(1+\gamma_5)Q|T_Q\ra 
\non \\  &=& -{1\over 2}\tilde B(a_q+b_q)(2m_{T_Q}),   \non \\
\la \Omega_Q|\bar Q^\alpha(1-\gamma_5)s^\beta\bar s^\beta(1+\gamma_5)
Q^\alpha|\Omega_Q\ra &=& -\tilde B\la \Omega_Q|\bar Q(1-\gamma_5)s\bar s(1+
\gamma_5)Q|\Omega_Q\ra   \non \\
&=& \tilde B(a_s-{5\over 3}b_s-{16\over 3}c_s)(2m_{\Omega_Q}).
\en

  In the nonrelativistic quark model (NQM), baryon matrix elements of 
four-quark operators are the same as that of (2.18) and (2.22) except for
the replacement:
\be
a_q\to |\psi_{Qq}(0)|^2=\int d^3r\,u_q^2(r)u_Q^2(r),~~~~~b_q\to 0,~~~~c_q\to 0.
\en
In general, the strength of destructive Pauli interference and $W$-exchange
is governed by $a_q+b_q$ in the bag model and $|\psi(0)|^2$ in the NQM. 
However, it is well known in hyperon decay that the bag model calculation
of $a_q+b_q$ gives a much smaller value than the  nonrelativistic estimate
of $|\psi(0)|^2$: $a_u+b_u\sim 3\times 10^{-3}{\rm GeV}^3$, while $|\psi(0)|
^2\sim 10^{-2}{\rm GeV}^3$. We shall see later that this also occurs in bottom
baryon decay. As pointed out in \cite{Cheng92}, naively one may be tempted
to conclude that the relativistic models are presumably more reliable. For
example, the lower component of the wavefunction is needed to reduce the 
NQM prediction $g_A={5\over 3}$ to the experimental value of 1.25. However,
the difference between $a_u+b_u$ and $|\psi(0)|^2$ is not simply attributed
to relativistic corrections; it arises essentially from the
distinction in the spatial scale of the wavefunction especially at the origin.
As a consequence, both models give a quite different quantitative 
description for processes sensitive to $|\psi(0)|^2$. It has been long
advocated in \cite{LaYaou} that a small value of $|\psi(0)|^2$ should be
discarded since a realistic potential that fits to the orbital-excitation
spectrum yields $\la\delta(\vec{r}_1-\vec{r}_2)\ra\sim 10^{-2}{\rm GeV}^3$.
Empirically, it also appears that the NQM works better for charmed baryon
decays \cite{Guberina,Cheng92}.

   In the following we will consider the NQM estimate of baryon
matrix elements. Consider $|\psi^{\Lambda_b}_{bq}(0)|^2$ as an example.
A straightforward calculation of hyperfine splitting between $\Sigma_b$
and $\Lambda_b$ yields \cite{Rujula}
\be
m_{\Sigma_b}-m_{\Lambda_b}=\,{16\pi\over 9}\,\alpha_s(m_b)\,{m_b-m_q\over
m_bm_q^2}\,|\psi_{bq}^{\Lambda_b}(0)|^2,
\en
where the equality $|\psi^{\Sigma_b}_{bq}(0)|^2=|\psi^{\Lambda_b}_{bq}(0)|^2$ 
has been assumed. The uncertainties in Eq.~(2.24) associated with 
$\alpha_s(m_b)$ and the constituent
quark mass $m_q$ can be reduced by introducing the $B$-meson wavefunction
at the origin squared $|\psi^B_{b\bar q}(0)|^2={1\over 12}f_B^2m_B$ which 
is related to the $B^*$ and $B$ mass difference by $m_{B^*}-m_B={32\over 9}
\pi\alpha_s(m_b)|\psi^B_{b\bar q}(0)|^2/(m_bm_q)$. Hence,
\be
|\psi^{\Lambda_b}_{bq}(0)|^2=\,{2m_q\over m_b-m_q}\,{m_{\Sigma_b}-
m_{\Lambda_b}\over m_{B^*}-m_B}\,|\psi^B_{b\bar q}(0)|^2.
\en 
Another method is proposed by Rosner \cite{Rosner} to consider the
hyperfine splittings of $\Sigma_b$ and $B$ separately so that
\be
|\psi^{\Lambda_b}_{bq}(0)|^2=|\psi^{\Sigma_b}_{bq}(0)|^2=\,{4\over 3}\,{m
_{\Sigma^*_b}-m_{\Sigma_b}\over m_{B^*}-m_B}\,|\psi^B_{b\bar q}(0)|^2.
\en
This method is supposed to be most reliable as $|\psi_{bq}(0)|^2$ thus
determined does not depend on $\alpha_s$ and $m_q$ directly.
Numerically, we find that (2.25) and (2.26) both give very similar results.
Defining the wavefunction ratio
\be
r=\left|{\psi^{\Lambda_b}_{bq}(0)\over \psi^B_{b\bar q}(0)}\right|^2,
\en
the baryon matrix elements in (2.18) and (2.22) can be recast to
\be
&& \la T_b|(\bar bb)(\bar qq)|T_b\ra =-\tilde B\la T_b|(\bar bq)(\bar qb)|T_b
\ra=\,{1\over 12}f_{B_q}^2m_{_{B_q}}r\tilde B(2m_{T_b}),   \non \\
&& \la T_b|\bar b(1-\gamma_5)q\bar q(1+\gamma_5)b|T_b\ra =\,{1\over 24}
f_{B_q}^2m_{_{B_q}}r(2m_{T_b}),   \non \\
&& \la T_b|\bar b^\alpha(1-\gamma_5)q^\beta\bar q^\beta(1+\gamma_5)b^\alpha
|T_b\ra =\,-{1\over 24}f_{B_q}^2m_{_{B_q}}r\tilde B(2m_{T_b}),   \non \\
&& \la \Omega_b|(\bar bb)(\bar ss)|\Omega_b\ra =-\tilde B\la \Omega_b|(\bar 
bs)(\bar sb)|\Omega_b\ra=\,{1\over 2}f_{B_q}^2m_{_{B_q}}r\tilde B(2m_{
\Omega_b}),   \non \\
&& \la \Omega_b|\bar b(1-\gamma_5)s\bar s(1+\gamma_5)b|\Omega_b\ra =\,-{1\over 
12}f_{B_q}^2m_{_{B_q}}r(2m_{\Omega_b}),   \non \\
&& \la \Omega_b|\bar b^\alpha(1-\gamma_5)s^\beta\bar s^\beta(1+\gamma_5)
b^\alpha|\Omega_b\ra =\,{1\over 12}f_{B_q}^2m_{_{B_q}}r\tilde B
(2m_{\Omega_b}),   
\en
where $f_{B_q}$ is the decay constant of the meson $\bar B_q$.

    To estimate $|\psi_{bq}(0)|^2$ and the parameter $r$ in the NQM, we find 
from (2.26)
\footnote{Our result for $r_{\Lambda_b}$ is the same as \cite{Rosner} but
different from \cite{Neubert97} in which $r_{\Lambda_b}$ is given by
${4\over 3}(m^2_{\Sigma_b^*}-m^2_{\Sigma_b})/(m^2_{B^*}-m^2_B)$.}
\be
r_{\Lambda_b}=\,{4\over 3}\,{m_{\Sigma_b^*}-m_{\Sigma_b}\over m_{B^*}-m_B},~~~
r_{\Xi_b}=\,{4\over 3}\,{m_{\Xi_b^*}-m_{\Xi'_b}\over m_{B^*}-m_B},~~~
r_{\Omega_b}=\,{4\over 3}\,{m_{\Omega_b^*}-m_{\Omega_b}\over m_{B^*}-m_B},
\en
and likewise for $r_{\Lambda_c},~r_{\Xi_c}$ and $r_{\Omega_c}$, where $\Xi'_{
b,c}$ denote spin-${1\over 2}$ sextets. Heavy baryon masses have been studied 
in \cite{Jenkins} in $1/m_Q$ and $1/N_c$ expansions within the HQET framework.
The chromomagnetic mass splittings for charmed baryons are given by
\cite{Jenkins}
\be
&& m_{\Sigma_c^*}-m_{\Sigma_c}=\,65.7\pm 2.3\,{\rm MeV},~~~m_{\Xi_c^*}-
m_{\Xi'_c}=\,63.2\pm 2.6\,{\rm MeV},   \non \\
&& m_{\Omega_c^*}-m_{\Omega_c}=\,60.6\pm 5.7\,{\rm 
MeV},
\en
where precise measurements of $\Sigma_c^*$ and $\Xi_c^*$ have been reported
by CLEO \cite{CLEO}. It is evident that the heavy-quark spin-violating mass 
relation \cite{Jenkins}
\be
(m_{\Sigma_c^*}-m_{\Sigma_c})+(m_{\Omega_c^*}-m_{\Omega_c})=2(m_{\Xi_c^*}
-m_{\Xi'_c})
\en
is very accurate. It follows that 
\be
m_{\Sigma_b^*}-m_{\Sigma_b}=\,\left({m_c\over m_b}\right)(m_{\Sigma_c^*}-m_{
\Sigma_c})=21.0\,{\rm MeV}
\en
for $m_b=5$ GeV, $m_c=1.6$ GeV (see below). This mass splitting is 
substantially smaller than the preliminary 
result $m_{\Sigma_b^*}-m_{\Sigma_b}=(56\pm 16){\rm MeV}$ reported by the 
DELPHI Collaboration \cite{DELPHI}. Since the measured mass difference 
of $\Sigma_c^*$ and $\Sigma_c$ is around 66 MeV [cf. (2.30)], a large 
hyperfine splitting of order 55 MeV for the $\Sigma_b$ baryon is very
unlikely. Likewise,
\be
m_{\Xi_b^*}-m_{\Xi'_b}=\,20.2\,{\rm MeV},~~~~m_{\Omega_b^*}-m_{\Omega_b}=\,
19.4\,{\rm MeV}.
\en
Because $\Delta m_B=m_{B^*}-m_B=45.7\pm 0.4$ MeV and $\Delta m_D=m_{D^*}-m_D
\cong 143$ MeV 
\cite{PDG} [note that $\Delta m_B$ and $\Delta m_D$ obey the same scaling 
relation as (2.32)], we find
\be
r_{\Lambda_c}\cong r_{\Lambda_b}=0.61,~~~
r_{\Xi_c}\cong r_{\Xi_b}=0.59,~~~
r_{\Omega_c}\cong r_{\Omega_b}=0.53,
\en
and
\be
&& |\psi^{\Lambda_b}_{bq}(0)|^2=0.87\times 10^{-2}{\rm GeV}^3,~~~
|\psi^{\Xi_b}_{bq}(0)|^2=0.84\times 10^{-2}{\rm GeV}^3,   \non \\
&& |\psi^{\Omega_b}_{bq}(0)|^2=0.81\times 10^{-2}{\rm GeV}^3,
\en
for $f_{B_q}=180$ MeV \cite{Sachrajda}. An estimate in the QCD sum 
rule analysis yields $r\simeq 0.1-0.3$ \cite{Col}.
Therefore, the NQM estimate of $|\psi_{bq}(0)|^2$ is 
indeed larger than the analogous bag model quantity: $a_q+b_q\sim 3\times
10^{-3}{\rm GeV}^3$. However, for the charmed baryon we obtain 
$|\psi^{\Lambda_c}_{cq}(0)|^2=3.8\times 10^{-3}{\rm GeV}^3$ for 
$f_D=200$ MeV \cite{Sachrajda}, which 
is smaller than those in bottom or hyperon decay. It seems that the smallness
of $|\psi^{\Lambda_c}_{cq}(0)|^2$ is ascribed to the assumption that the 
$D$ meson wavefunction at the origin squared $|\psi^D_{c\bar q}(0)|^2$ is 
given by ${1\over 12}f_D^2m_D$. We will come back to this point in Sec.~IV.
By comparing (2.28) with (2.18) we see that $r$ is of order 0.20 in the 
bag model.

   Finally we are ready to estimate the HQET parameter $\lambda_2^{\rm
baryon}$ [see Eq.~(2.12)]. Using the baryon masses \cite{Jenkins}
\be
m_{\Sigma_c}=2452.9\,{\rm MeV},~~~~m_{\Xi'_c}=2580.8\,{\rm MeV},~~~~m_{
\Omega_c}=2699.9\,{\rm MeV},   \non \\
m_{\Sigma_b}=5824.2\,{\rm MeV},~~~~m_{\Xi'_b}=5950.9\,{\rm MeV},~~~~m_{
\Omega_b}=6068.7\,{\rm MeV},   
\en
and (2.30)-(2.33) we find
\be
\lambda_2^{\rm baryon}=\cases{ 0.055\,{\rm GeV}^2 & for~charmed baryons; \cr
0.041\,{\rm GeV}^2 & for~$\Sigma_b$; \cr  0.040\,{\rm GeV}^2 & for~$\Xi'_b$; 
\cr  0.039\,{\rm GeV}^2 & for~$\Omega_b$. \cr  }
\en
It is interesting to note that the large-$N_c$ relation \cite{Jenkins}
\be
\lambda_2^{\rm meson}\sim N_c\lambda_2^{\rm baryon}
\en
is fairly satisfied especially for bottom hadrons.

\section{Lifetimes of Bottom Hadrons}
   Using the formulism described in the last section, semileptonic and 
nonleptonic widths are calculated in this section.
    We shall first try to fix the heavy quark pole mass from the measured 
inclusive semileptonic decay rate. The semileptonic width of the $B$
meson given by (2.5)
\be
&& \Gamma_{\rm SL}(B\to Xe\bar\nu)=\,{G_F^2m_b^5\over 192\pi^3}|V_{cb}|^2 
\eta(x,0,0)
\left\{ I_0(x,0,0){\la \bar B|\bar bb|\bar B\ra\over 2m_B}-{1\over m_b^2}I_1
(x,0,0){\la\bar B|\bar b\sigma\cdot Gb|\bar B\ra\over 2m_B}\right\} 
\en
has the salient feature that empirically $\Gamma_{\rm SL}(B)$ is very 
insensitive to the choice of $m_b$ as long as $m_b-m_c$, which is free of
renormalon ambiguity, is fixed according to Eq.~(2.14). Hence, we may use
the measured $\Gamma_{\rm SL}(D)$ to fix $m_c$ to be 1.6 GeV (see Sec.~IV
below) which in turn implies $m_b=5$ GeV, in excellent agreement with the pole
mass determined from lattice QCD: $m_b=5.0\pm 0.2$ GeV \cite{Davis}.
Since $x=(m_c/m_b)^2=0.1024$, 
the phase-space factors $I_i$ in (2.2) and (2.3) read
\be
&& I_0(x,0,0)=0.476,~~~I_0(x,x,0)=0.147,~~~I_1(x,0,0)=0.649,   \non \\
&&I_1(x,x,0)=0.328,~~~I_2(x,0,0)=0.723,~~~I_2(x,x,0)=0.220.
\en
From (3.1) we obtain
\be
&& \Gamma(B\to Xe\bar\nu) =4.44\times 10^{-14}{\rm GeV}, \non \\
&& \Gamma_{\rm SL}(B)=2.24\,\Gamma(B\to Xe\bar \nu)=9.95\times 
10^{-14}{\rm GeV},
\en
for $|V_{cb}|=0.039$, where uses of Eqs.~(2.8), (2.9), (2.12) and (2.13) 
have been made, for example,
\be
{\la \bar B|\bar b\sigma\cdot G b|\bar B\ra\over 2m_B}=\,6\lambda^{\rm 
meson}_2=\,0.72\,{\rm GeV}^2.
\en
Since the phase space for the $\tau$ semileptonic decay mode relative to 
that of the $e$ mode is $0.24:1$, this accounts for the factor 2.24 in 
(3.3). The result (3.3) agrees very well with experiment \cite{PDG}
\be
\Gamma(B^-/B^0\,{\rm admixture}\to Xe\bar\nu)=(4.31\pm 0.17)\times 10^{-14}
{\rm GeV}.
\en
Likewise, we find for bottom-baryon semileptonic decays
\be
&& \Gamma(\Lambda_b\to Xe\bar\nu)=\Gamma(\Xi_b\to Xe\bar\nu)=\,4.59\times
10^{-14}{\rm GeV},   \non \\
&& \Gamma(\Omega_b\to Xe\bar\nu)=\,4.53\times 10^{-14}{\rm GeV},
\en
and hence
\be
&& \Gamma_{\rm SL}(\Lambda_b)=\Gamma_{\rm SL}(\Xi_b)=2.24\,\Gamma(\Lambda_b\to
Xe\bar\nu)=\,1.027\times 10^{-13}{\rm GeV},   \non \\
&& \Gamma_{\rm SL}(\Omega_b)=2.24\,\Gamma(\Omega_b\to
Xe\bar\nu)=\,1.014\times 10^{-13}{\rm GeV}.
\en
Note that the tiny difference between $\Gamma_{\rm SL}(\Lambda_b)$ and
$\Gamma_{\rm SL}(\Omega_b)$ arises from the fact that the chromomagnetic
operator contributes to the matrix element of $\Omega_b$ but not to 
$\Lambda_b$ (or $\Xi_b$) as the light degrees of freedom in the latter are 
spinless; that is,
\be
{\la\Lambda_b|\bar b\sigma\cdot Gb|\Lambda_b\ra\over 2m_{\Lambda_b}}=0,~~~~
{\la\Omega_b|\bar b\sigma\cdot Gb|\Omega_b\ra\over 2m_{\Omega_b}}=8
\lambda_2^{\rm baryon}=0.31\,{\rm GeV}^2.
\en

   To compute the nonleptonic decay rate we apply the Wilson coefficient
functions
\be
c_1(\mu)=1.14,~~~~~c_2(\mu)=-0.31,
\en
which are evaluated at $\mu=4.4$ GeV to the leading logarithmic approximation
(see Table XIII of \cite{Buras}). From 
Eq.~(2.1) the nonleptonic widths of bottom baryons arising from $b$ 
quark decay are found to be
\be
&& \Gamma^{\rm dec}(B)=2.216\times 10^{-13}{\rm GeV},~~~~~~\Gamma^{\rm dec}
(\Omega_b)=2.217\times 10^{-13}{\rm GeV},   \non \\
&& \Gamma^{\rm dec}(\Lambda_b)=\Gamma^{\rm dec}(\Xi_b)=2.220
\times 10^{-13}{\rm GeV}.
\en
We see that the $b$ quark decay contribution $\Gamma^{\rm dec}$ is very 
similar for
bottom hadrons even though the chromomagnetic mass splitting is different
among them. Therefore, to ${\cal O}(1/m_b^3)$ we obtain
\be
{\tau(\Lambda_b)\over\tau(B_d)}\cong{\tau(\Xi_b)\over\tau(B_d)}\cong
{\tau(\Omega_b)\over\tau(B_d)}=0.99+{\cal O}(1/m_b^3).
\en

   We next turn to the nonspectator effects of order $1/m_b^3$.
The Pauli interference in inclusive nonleptonic $B^-$ decay and the 
$W$-exchange contribution to $B_d$ can be evaluated from the first and second
terms in (2.4):
\be
\Gamma^{\rm ann}(B_d) &=& -\Gamma_0\,\eta_{\rm nspec}\Bigg\{(1-x)^2(1+{1
\over 2}x)\Big[({1\over 3}c_1^2+2c_1c_2+N_cc_2^2)B_1+2c_1^2\ep_1\Big]  \non \\
&& -(1-x)^2(1+2x)\Big[({1\over 3}c_1^2+2c_1c_2+N_cc_2^2)B_2+2c_1^2\ep_2\Big]
\Bigg\}, \non \\
\Gamma^{\rm int}_-(B^-) &=& \Gamma_0\,\eta_{\rm nspec}(1-x)^2\left[(c_1^2+
c_2^2)(B_1+6\ep_1)+6c_1c_2B_1\right],
\en
with \cite{Neubert97}
\be
\Gamma_0={G_F^2m_b^5\over 192\pi^3}|V_{cb}V_{ud}|^2,~~~\eta_{\rm nspec}=16
\pi^2{f_B^2m_B\over m_b^3},
\en
where we have applied Eqs.~(2.15), (2.16) and neglected Cabibbo-suppressed
$W$-exchange contribution to $B_d$. As stressed in \cite{Neubert97},
the coefficients of $B_i$ in (3.12) are one to two orders of magnitude
smaller than that of $\ep_i$. Therefore, the contributions of $B_i$ can be
safely neglected at least in $\Gamma^{\rm ann}(B_d)$. Numerically,
\be
\Gamma^{\rm ann}(B_d) &=& (-0.491\ep_1+0.563\ep_2)\times 10^{-13}
{\rm GeV},   \non \\
\Gamma^{\rm int}_-(B^-) &=& (-0.130B_1+1.505\ep_1)\times 10^{-13}{\rm GeV}.
\en
Beyond the factorization approximation, $\ep_i$ may receive nonfactorizable
contributions. A QCD sum rule estimate gives $\ep_1\approx -0.15$ and
$\ep_2\approx 0$ \cite{Chernyak}. This implies a constructive $W$-exchange
to $B_d$ and a destructive Pauli interference to $B^-$. 

   As for the nonspectator effects in nonleptonic decays of bottom baryons 
we obtain from (2.4) that
\be
\Gamma^{\rm ann}(\Lambda_b) &=& \Gamma_0\,\eta_{\rm nspec}\,r(1-x)^2\Big
(\tilde B(c_1^2+c_2^2)-2c_1c_2\Big),   \non \\
\Gamma^{\rm int}_-(\Lambda_b) &=& -{1\over 4}\Gamma_0\,\eta_{\rm nspec}\,r
\left[(1-x)^2(1+x)+\left|{V_{cd}\over V_{ud}}\right|^2\sqrt{1-4x}\,\right]
\Big(\tilde Bc_1^2-2c_1c_2-N_cc_2^2\Big),   \non \\
\Gamma^{\rm ann}(\Xi_b^0) &=& \Gamma^{\rm ann}(\Lambda_b),~~~~~~~\Gamma^{\rm
int}_-(\Xi_b^0)=\,\Gamma^{\rm int}_-(\Lambda_b),   \non \\
\Gamma^{\rm int}_-(\Xi_b^-) &=& -{1\over 4}\Gamma_0\,\eta_{\rm nspec}\,r\Big
[(1-x)^2(1+x)+\sqrt{1-4x}\,\Big]\Big(\tilde{B}c_1^2-2c_1c_2-N_cc_2^2\Big), 
\non \\
\Gamma^{\rm int}_-(\Omega_b^-) &=& -{1\over 6}\Gamma_0\,\eta_{\rm nspec}\,r
\sqrt{1-4x}\,(5-8x)\Big(\tilde{B}c_1^2-
2c_1c_2-N_cc_2^2\Big),
\en
where use has been made of (2.28). Note that there is no $W$-exchange
contribution to the $\Xi_b^-$ and $\Omega_b$
and that there are two Cabibbo-allowed Pauli interference terms in $\Xi_b^-$ 
decay, and one Cabibbo-allowed as well as one Cabibbo-suppressed 
interferences in $\Lambda_b$ decay. It is easily seen that under the 
valence-quark approximation i.e. $\tilde B=1$, the $W$-exchange contribution 
$\Gamma^{\rm ann}$
is proportional to $c_-=(c_1-c_2)/2$ as the four-quark operator $O_+=(\bar 
q_1q_2)(\bar q_3q_4)+(\bar q_1q_4)(\bar q_3q_2)$ is symmetric in color
indices whereas the color wavefunction for a baryon is totally antisymmetric.
Writing
\be
\Gamma_{\rm NL}=\Gamma^{\rm dec}+\Gamma^{\rm ann}+\Gamma^{\rm int}_-,
\en
the numerical results for nonleptonic inclusive decay rates are
\be
\Gamma_{\rm NL}(\Lambda_b) &=& \Big[\,2.220+(0.042+0.058\tilde B)r\Big]\times 
10^{-13}{\rm GeV},    \non \\
\Gamma_{\rm NL}(\Xi_b^0) &=& \Big[\,2.220+(0.043+0.066\tilde B)r\Big]\times 
10^{-13}{\rm GeV},    \non \\
\Gamma_{\rm NL}(\Xi_b^-) &=& \Big[\,2.220-(0.037+0.114\tilde B)r\Big]\times 
10^{-13}{\rm GeV},    \non \\
\Gamma_{\rm NL}(\Omega_b) &=& \Big[\,2.217-(0.043+0.133\tilde B)r\Big]\times 
10^{-13}{\rm GeV},    
\en
where for later convenience we have normalized the parameter $r$ in (3.17) 
to $r_{\Lambda_b}$ [see (2.34)]; that is, we have taken into account SU(3)
breaking effect for $r$. Note that $\ep_i,~B_i$ in (3.14) and $\tilde
B,~r$ in (3.17) are all renormalized at $\mu=4.4$ GeV.

   Before proceeding, it is worth emphasizing the difference between the 
$W$-exchange contributions in the inclusive nonleptonic decays of the $B$ 
meson and the bottom baryon. It is conventionally argued that $W$-exchange 
in heavy meson decay is suppressed by helicity and color mismatch. For
example, $W$-exchange in $B$ decay is helicity suppressed by a factor of 
$16\pi^2(f_B/m_B)^2$ relative to the heavy quark decay amplitude.
\footnote{It had been claimed that soft gluon emission from the 
initial quark line or soft gluon content in the initial wavefunction can
vitiate both helicity and color suppression \cite{Bander}. The net effect is 
that the factor $f_B/m_B$ is effectively replaced by $f_B/m_q$, where 
$m_q$ is the constituent quark mass of the antiquark in the $\bar B$ 
meson \cite{BU}. As a consequence, contributions of $W$-exchange will 
exhibit powerlike $(m_B/m_q)^2$
enhancement and this renders the treatment of the heavy quark expansion 
for $W$-exchange invalid. This issue was resolved by Bigi and Uraltsev
\cite{BU} who showed that such powerlike enhancement does not arise for
fully inclusive transitions and the soft gluon effect merely amounts to
renormalizing the coefficients of 4-quark operators.}
By contrast, $W$-exchange in baryon decay is neither helicity nor color
suppressed. The diquark $Qq$ system in the heavy baryon can have a spin 0
configuration and the decay of a spin 0 (not spin 1 !) state into two quarks 
is not subject to helicity suppression.

    Since $\tilde B$ is of order unity and $r\sim 0.60$, it is evident from
(3.17) and (3.10) that the bottom baryon lifetimes follow the pattern (see
also Table I below)
\be
\tau(\Omega_b^-)\simeq\tau(\Xi_b^-)>\tau(\Lambda_b^0)\simeq\tau(\Xi_b^0).
\en
This pattern originates from the fact that while $\Lambda_b,~\Xi_b^0,~\Xi_b
^-,~\Omega_b$ all receive contributions from destructive Pauli interference,
only $\Lambda_b$ and $\Xi_b^0$ have $W$-exchange and that $\Gamma^{\rm int}
_-$ is most large in $\Omega_b$ due to the presence of two valence $s$ quarks 
in its quark content. We shall see shortly that this lifetime pattern is 
dramatically modified when the $b$ quark mass is replaced by the bottom 
baryon mass in nonleptonic widths.

   It follows from (3.3), (3.7), (3.10), (3.14) and (3.17) that
\be
{\tau(\Lambda_b)\over\tau(B_d)} =\,0.99-0.15\ep_1+0.17\ep_2-(0.013+
0.018\tilde B)r,
\en
which is a model-independent result. This is consistent with the
result (1.7)
obtained in \cite{Neubert97} with $\ep_i,~\tilde B,~r$ renormalized at
$\mu=4.85$ GeV and with $f_B=200$ MeV.
As stated in the Introduction, $\ep_1$ and $\ep_2$ obey the constraint
$\ep_1\approx 0.3\,\ep_2$, then it is quite difficult, if not impossible, to 
accommodate the experimental value (1.5) for $\tau(\Lambda_b)/\tau(B_d)$
without invoking too large value of $r$ and/or $\tilde B$. 
We will argue below that the contribution of $-0.15\ep_1+\cdots-0.018\tilde
Br$ in (3.19) is at most of order 6\%. We hasten to remark that the current
CDF result (1.6) for the lifetime ratio is consistent with the theoretical
prediction (see, however, a comment after Eq.~(3.22)).

 Irrespective of the above-mentioned lifetime ratio problem, there exists 
another serious difficulty, namely the predicted absolute decay width of the 
$B$ or $\Lambda_b$ hadron based on the heavy quark expansion [see (3.3), 
(3.7), (3.10) and (3.17)] is too small 
compared to the experimental values \cite{LEP}:
\be
\Gamma(B_d) &=& \left(4.246^{+0.094}_{-0.125}\right)\times
10^{-13}{\rm GeV},~~~~\tau(B_d)=\,(1.55\pm 0.04)\,ps,   \non \\
\Gamma(B^-) &=& \left(3.965^{+0.098}_{-0.093}\right)\times
10^{-13}{\rm GeV},~~~~\tau(B^-)=\,(1.66\pm 0.04)\,ps,   \non \\
\Gamma(\Lambda_b) &=& \left(5.351^{+0.422}_{-0.365}\right)\times
10^{-13}{\rm GeV},~~~~\tau(\Lambda_b)=\,(1.23\pm 0.09)\,ps.
\en
Obviously, even if the destructive contribution $\Gamma^{\rm int}_-(B^-)$
is not taken into account, the result $\Gamma^{\rm dec}(B)+\Gamma_{\rm SL}(B)
=3.211\times 10^{-13}$ GeV is too small by about 20\% to account for the 
observed decay rate of $B^-$. 
\footnote{The problem with the absolute total decay width $\Gamma(B)$ of the 
$B$ meson is
intimately related to the problem with the $B$-meson semileptonic branching 
ratio $B_{\rm SL}$. The theoretical prediction for $B_{\rm SL}$ is in general 
above 12.5\% \cite{BigiSL}, while experimentally $B_{\rm SL}=(10.23\pm 
0.39)\%$ \cite{Richman}. In our case we obtain $B_{\rm SL}\gtrsim 13.8\%$.
Several scenarios have been put forward in the
past to resolve the discrepancy between theory and experiment for $B_{\rm
SL}$ or $\Gamma(B)$. Here we mention two of the possibilities. (i) Since
the theoretical results depend on the scale $\mu$ to renormalize 
$\alpha_s(\mu)$ and the Wilson coefficients $c_{1,2}(\mu)$, one may
choose a low renormalization scale, $\mu/m_b\sim 0.3-0.5$, to accommodate
the data \cite{Neubert97}. Local duality holds in this scenario. (ii)
Next-to-leading order QCD radiative corrections to nonleptonic decay will 
increase the rate for $b\to c\bar cs$ substantially and decrease $B_{\rm SL}$
\cite{Bagan,Voloshin95}. Using the result of \cite{Bagan}, we find that the
QCD effect will bring $B_{\rm SL}$ down by 1\% and hence $B_{\rm SL}\gtrsim
12.7\%$. It was suggested in \cite{Palmer,Falk} that a failure
of local duality in the $b\to c\bar cs$ channel, which has smaller energy
release than that in $b\to c\bar ud$, will further enhance $\Gamma(B)$ and
suppress $B_{\rm SL}$. However, this explanation encounters a problem: The 
charm counting $n_c$ will increase and become as large as 1.30 \cite{Bagan},
which is too large compared to the experimental value $n_c=1.12\pm 0.05$
\cite{Richman}. One way out of this difficulty for $n_c$ is proposed in
\cite{Dunietz} that a sizeable fraction of $b\to c\bar cs$ transitions
can be seen as charmless $b\to s$ processes. In the present paper we will
not pursue any of the aforementioned possibilities as
none of them can explain the lifetime difference
between $\Lambda_b$ and $B_d$. The recipe we are going to discuss below
[see (3.23)] will solve all the problems with $B_{\rm SL},~\Gamma(B),~n_c$ and
$\tau(\Lambda_b)/\tau(B_d)$.}
To compute the decay widths of bottom baryons, 
we have to specify the values of $\tilde B$ and $r$. Since $\tilde B=1$ in the
valence-quark approximation and since the wavefunction squared ratio $r$
is evaluated using the quark model, it is reasonable to assume that the NQM
and the valence-quark approximation are most reliable when the baryon matrix
elements are evaluated at a typical hadronic scale $\mu_{\rm had}$. As
shown in \cite{Neubert97}, the parameters $\tilde B$ and $r$ renormalized
at two different scales are related via the renormalization group equation
to be
\be
&& \tilde B(\mu)r(\mu) =\, \tilde B(\mu_{\rm had})r(\mu_{\rm had}),   \non \\
&& \tilde B(\mu) =\, {\tilde{B}(\mu_{\rm had})\over \kappa+{1\over N_c}(\kappa
-1)\tilde B(\mu_{\rm had}) }\,,
\en
with
\be
\kappa=\left({\alpha_s(\mu_{\rm had})\over \alpha_s(\mu)}\right)^{3N_c/2
\beta_0}=\sqrt{\alpha_s(\mu_{\rm had})\over \alpha_s(\mu)}
\en
and $\beta_0={11\over 3}N_c-{2\over 3}n_f$. Choosing $\alpha_s(\mu_{\rm had})
=0.5$ and $\mu=4.4$ GeV, we obtain $\tilde{B}(\mu)=0.59\tilde B(\mu_{\rm 
had})\simeq 0.59$ and $r(\mu)\simeq 1.7\,r(\mu_{\rm had})$. Using $r(\mu_{\rm 
had})=0.61$ [see (2.34)], the calculated decay rates of bottom baryons 
are summarized in Table I. It is evident that the predicted $\Lambda_b$ 
lifetime is too large by 8 standard deviations. Note that while the CDF 
measurement
(1.6) for the lifetime ratio $\tau(\Lambda_b)/\tau(B_d)$ can be easily
accommodated in theory, it is still difficult to explain the absolute lifetime
$\tau(\Lambda_b)=(1.32\pm 0.15\pm 0.07)ps$ measured by CDF \cite{Tseng}.

\vskip 0.4cm
{{\small Table I. Various contributions to the decay rates (in units of
$10^{-13}$ GeV) of bottom baryons.}}
{
\begin{center}
\begin{tabular}{|c|c c c c c c c|} \hline
 & $\Gamma^{\rm dec}$ & $\Gamma^{\rm ann}$ & $\Gamma^{\rm int}_-$ & 
$\Gamma_{\rm SL}$ & $\Gamma^{\rm tot}$ & $\tau(10^{-12}s)$~ & ~
$\tau_{\rm expt}(10^{-12}s)$~ \\  \hline 
$\Lambda_b^0$~ & ~2.220~ & ~0.145~ & ~$-0.064$~ & ~1.027~ & ~3.327~ & ~1.98~ 
& ~$1.23\pm 0.09$~   \\
$\Xi_b^0$ & 2.220 & 0.138 & $-0.051$ & 1.027 & 3.334 & 1.97 &   \\
$\Xi_b^-$ & 2.220 & & $-0.110$ & 1.027 & 3.137 & 2.10 &  \\
$\Omega_b^-$ & 2.217 & & $-0.127$ & 1.014 & 3.104 & 2.12 &   \\
\hline
\end{tabular}
\end{center} }
\vskip 0.4cm

    It has been advocated in \cite{Altarelli} that, unlike the 
semileptonic inclusive case, since OPE cannot be rigorously justified for
nonleptonic inclusive decays, the failure of explaining the observed
lifetime ratio $\tau(\Lambda_b)/\tau(B_d)$ implies that the
assumption of local duality is not correct for nonleptonic inclusive widths.
It is further suggested in \cite{Altarelli} that corrections of order $1/m_Q$
should be present and this amounts to replacing the heavy quark mass 
by the mass of the decaying hadron in the $m^5_Q$ factor in front of all
nonleptonic widths. In the following we shall see that the ansatz
\be
\Gamma_{\rm NL} \longrightarrow \Gamma_{\rm NL}\left({m_{H_b}\over m_b}\right
)^5
\en
will not only solve the short $\Lambda_b$ lifetime problem but also provide 
the correct absolute decay rates for bottom hadrons.

   Employing the hadron masses
\be
&& m_{B_d}=\,5279.2\pm 1.8\,{\rm MeV}~\cite{PDG},~~~~m_{B^-}=\,5278.9\pm 1.8\,
{\rm MeV}~\cite{PDG},   \non \\
&& m_{\Lambda_b}=\,5621\pm 5\,{\rm MeV}~\cite{CDF},
\en
we obtain
\be
\Gamma_{\rm tot}(\Lambda_b) &=& \Big[\,3.986+(0.075+0.105\tilde B)r\Big]
\times 10^{-13}{\rm GeV}+\Gamma_{\rm SL}(\Lambda_b),   \non \\
\Gamma_{\rm tot}(B_d) &=& \Big[\,2.908+(-0.644\ep_1+0.739\ep_2)\Big]
\times 10^{-13}{\rm GeV}+\Gamma_{\rm SL}(B),  \non \\ 
\Gamma_{\rm tot}(B^-) &=& \Big[\,2.907+(-0.171B_1+1.974\ep_1)\Big]
\times 10^{-13}{\rm GeV}+\Gamma_{\rm SL}(B),   
\en
with $\Gamma_{\rm SL}(\Lambda_b)$ and $\Gamma_{\rm SL}(B)$ being given 
by (3.3) and (3.7), respectively. Consequently,
\be
{\tau(\Lambda_b)\over\tau(B_d)} =\,0.78-0.13\ep_1+0.15\ep_2-(0.015+
0.021\tilde B)r.
\en
Comparing this with Eq.~(3.19) we see that the main effect of including 
linear $1/m_b$
corrections is to shift the central value of the lifetime ratio from 0.99
to 0.78. Moreover, the experimental value $\tau(\Lambda_b)/\tau(B_d)=0.79\pm
0.06$ \cite{LEP} indicates that the remaining contribution 
$-0.13\ep_1+\cdots$ in (3.26) is
at most $\pm 6\%$. It is also evident from (3.25) that the discrepancy
between theory and experiment for the absolute decay width of $B$ mesons
is greatly improved. 

   The most dramatic effect due to the ansatz (3.23) occurs in the lifetime
pattern of bottom baryons.
Employing the bottom-baryon masses (2.36), (3.24) and $m_{\Xi_b}=5803.7
\pm 7.1$ MeV, \footnote{We have used the CDF mass of the $\Lambda_b$ 
[see (3.24)] to update the $\Xi_b$ mass prediction given in \cite{Jenkins}.}
some large enhancement to various nonleptonic contributions to the decay 
widths of bottom baryons is shown in Table II.
We see that the improved $\Lambda_b$ lifetime is in agreement with
experiment and the new hierarchy of bottom-baryon lifetimes emerges as
\be
\tau(\Lambda_b^0)>\tau(\Xi_b^-)>\tau(\Xi_b^0)>\tau(\Omega_b^-),
\en
which is drastically different from the previous one: The longest-lived
$\Omega_b$ among bottom baryons in the conventional OPE now 
becomes 
shortest-lived. Needless to say, it is of great importance to measure the
hierarchy of bottom-baryon lifetimes in order to test the ansatz (3.23).
The branching ratios of semileptonic inclusive decays are calculated
from Table II to be:
\be
&& {\cal B}(\Lambda_b\to Xe\bar\nu)=\,8.9\%,~~~~~{\cal B}(\Xi_b^0\to Xe\bar
\nu)=\,7.8\%,    \non \\
&& {\cal B}(\Xi_b^-\to Xe\bar\nu)=\,8.4\%,~~~~~{\cal B}(\Omega_b\to Xe\bar
\nu)=\,6.9\%.
\en

    Since serious and precise measurements of the hierarchy of lifetimes of 
bottom baryons may not be available in the very near future,\footnote{ The
current LEP results for the lifetime of $\Xi_b$ are $(1.35^
{+0.37+0.15}_{-0.28-0.17})\,ps$ by ALEPH \cite{ALEPH} and $(1.5^{+0.7}_{-0.4}
\pm 0.3)\,ps$ by DELPHI \cite{DELPHI1}. The average is
$\tau(\Xi_b)=(1.39^{+0.34}_{-0.28})\,ps$. Evidently, the uncertainty is 
still too large to have a meaningful test on the prediction (3.27).}
it is thus important to carry out
more precise measurement of the $B_s$ lifetime. An application of the 
prescription (3.23) will modify the prediction \cite{Neubert97}
\be
{\tau(B_s)\over \tau(B_d)}=\,1\pm{\cal O}(1\%)
\en
to \cite{Altarelli}
\be
{\tau(B_s)\over \tau(B_d)}=\,0.938
\en
for the average $B_s$ lifetime. The current world average is $\tau(B_s)/\tau(
B_d)=0.98\pm 0.05$ \cite{LEP}.

\vskip 0.4cm
{{\small Table II. Various contributions to the decay rates (in units of
$10^{-13}$ GeV) of bottom baryons. The ansatz (3.23) has been applied to
enhance the nonleptonic $b$ quark decay and nonspectator effects.}}
{
\begin{center}
\begin{tabular}{|c|c c c c c c c|} \hline
 & $\Gamma^{\rm dec}$ & $\Gamma^{\rm ann}$ & $\Gamma^{\rm int}_-$ & 
$\Gamma_{\rm SL}$ & $\Gamma^{\rm tot}$ & $\tau(10^{-12}s)$~ & ~
$\tau_{\rm expt}(10^{-12}s)$~ \\  \hline
$\Lambda_b^0$~ & ~3.986~ & ~0.260~ & ~$-0.116$~ & ~1.027~ & ~5.157~ & ~1.28~ 
& ~$1.23\pm 0.09$~   \\
$\Xi_b^0$ & 4.678 & 0.290 & $-0.107$ & 1.027 & 5.888 & 1.12 &   \\
$\Xi_b^-$ & 4.678 & & $-0.231$ & 1.027 & 5.474 & 1.20 &  \\
$\Omega_b^-$ & 5.840 & & $-0.335$ & 1.014 & 6.519 & 1.01 &   \\
\hline
\end{tabular}
\end{center} }
\vskip 0.4cm

\section{Lifetimes Of Charmed Baryons}
  In Sec.~III we see that a replacement of the heavy quark mass with the 
decaying hadron mass in the $m_Q^5$ factor in front of nonleptonic widths
provides a much better description of the lifetimes of the $\Lambda_b$ baryon
and $B$ mesons. It is claimed in \cite{Altarelli} that a much better fit to
the charmed hadron lifetimes is also achieved if $\Gamma_{\rm NL}$ for 
charm decay approximately scales with the fifth power of charmed hadron
masses, apart from corrections of order $1/m_c^2$. We will carefully examine 
the applicability of this recipe in this section. For a theoretical
overview of charmed baryon lifetimes, the reader is referred to the review
of Blok and Shifman \cite{BS94}.

   We begin with the semileptonic inclusive decay of the $D$ meson:
\be
\Gamma(D\to Xe\bar\nu)=\,{G_F^2m_c^5\over 192\pi^3}|V_{cs}|^2 \eta(x,0,0)
\left\{ I_0(x,0,0){\la D|\bar cc|D\ra\over 2m_D}-{1\over m_c^2}I_1(x,0,0)
{\la D|\bar c\sigma\cdot Gc|D\ra\over 2m_D}\right\}.  
\en
We find that the
experimental values for $D^+$ and $D^0$ semileptonic widths \cite{PDG} can be
fitted by the quark pole mass $m_c=1.6$ GeV. Taking $m_s=170$ MeV, we 
then have $x=(m_s/m_c)^2=0.0113$ and
\be
I_0(x,0,0)=0.9166,~~~~I_1(x,0,0)=0.9556,~~~~I_2(x,0,0)=0.9665
\en
for charm decay. Repeating the same exercise for charmed baryons,
we obtain the charmed-baryon semileptonic decay rates
\be
&& \Gamma(\Lambda_c\to Xe\bar\nu)=\Gamma(\Xi_c\to Xe\bar\nu)=
\,1.533\times 10^{-13}{\rm GeV},   \non \\
&& \Gamma(\Omega_c\to Xe\bar\nu)=\,1.308\times 10^{-13}{\rm GeV},
\en
which are larger than that of the $D$ meson:
\be
\Gamma(D\to Xe\bar\nu)=\,1.090\times 10^{-13}{\rm GeV}.
\en
The prediction (4.3) for the $\Lambda_c$ baryon is in good agreement with
experiment
\be
\Gamma(\Lambda_c\to Xe\bar\nu)_{\rm expt}=\,(1.438\pm 0.543)\times 
10^{-13}{\rm GeV}.
\en
For charmed baryons $\Xi_c$ and $\Omega_c$, there is an additional
contribution to the semileptonic width coming from the Pauli 
interference of the $s$ quark \cite{Voloshin}. From (2.5) we obtain
\be
\Gamma^{\rm int}(\Xi_c\to Xe\bar\nu) &=& {1\over 4}\Gamma'_0\,\eta_{\rm 
nspec}\,r_{\Xi_c}(1-x^2)(1+x),   \non \\
\Gamma^{\rm int}(\Omega_c\to Xe\bar\nu) &=& {1\over 6}\Gamma'_0\,\eta_{\rm 
nspec}\,r_{\Omega_c}(1-x^2)(5+x),   
\en
where we have applied (2.28) for charmed baryon matrix elements,
$\Gamma'_0=\Gamma_0/|V_{ud}|^2$ and
\be
\Gamma_0={G_F^2m_c^5\over 192\pi^3}|V_{cs}V_{ud}|^2,~~~\eta_{\rm nspec}=16
\pi^2{f_D^2m_D\over m_c^3}.
\en
We shall see later that, depending on the parameter $r$, the nonspectator 
effect
in semileptonic decay of $\Xi_c$ and $\Omega_c$ can be very significant, in
particular for the latter.

    We now turn to the nonleptonic inclusive decays of charmed hadrons. It 
is well known that the longer lifetime of $D^+$ relative to $D^0$ comes 
mainly from the destructive Pauli interference in $D^+$ decay 
\cite{Guberina79,Bilic}. However, it 
is also known that, depending on the parameters $B_1$ and especially
$\ep_1$, the Pauli interference $\Gamma^{\rm int}_-(D^+)$ in analog to
$\Gamma^{\rm int}_-(B^-)$ given by (3.12) can be easily overestimated and may 
even overcome the $c$ quark decay rate $\Gamma^{\rm dec}$ so that the
resulting nonleptonic width becomes negative ! This certainly does not make
sense. It has been discussed in great length by Chernyak \cite{Chernyak} as 
how to circumvent the difficulty with the lifetime of $D^+$. We shall not 
address this issue in the present work and instead focus on the lifetimes of
charmed baryons. Our purpose is to apply the ansatz similar to (3.23) and see
if a better description of charmed baryon lifetimes can be achieved.

    In addition to the destructive Pauli interference $\Gamma^{\rm int}_-$,
there exists another Pauli interference term $\Gamma^{\rm int}_+$ in charmed
baryon decay which arises from the constructive interference between the $s$
quark produced in the $c$ quark decay and the spectator $s$ quark in the
charmed baryon. Since the expressions of $\Gamma^{\rm ann}$ and $\Gamma^{\rm
int}_-$ for charmed baryons are similar to (3.15) for bottom baryon decays,
here we will only write down the expressions for $\Gamma^{\rm int}_+$
described by the last term in Eq.~(2.4):
\be
\Gamma^{\rm int}_+(\Xi_c) &=& -{1\over 4}\Gamma_0\,\eta_{\rm nspec}\,r_{
\Xi_c}(1-x^2)(1+x)\Big(\tilde Bc_2^2-2c_1c_2-N_cc_1^2\Big),   \non \\
\Gamma^{\rm int}_+(\Omega_c) &=& -{1\over 6}\Gamma_0\,\eta_{\rm nspec}\,r_{
\Omega_c}(1-x^2)(5+x)\Big(\tilde Bc_2^2-2c_1c_2-N_cc_1^2\Big).   
\en
It is easily seen that (4.8) is reduced to (4.6) when $c_1=1,~c_2=0,~N_c=1$
and $V_{ud}=1$.
The $\Xi_c^+$ and $\Omega_c$ baryons also receive contributions from
Cabibbo-suppressed $W$-exchange:
\be
\Gamma^{\rm ann}(\Xi_c^+)&=& |V_{us}/V_{ud}|^2\,\Gamma_0\,\eta_{\rm nspec}\,
r_{\Xi_c}(1-x^2)\Big(\tilde B(c_1^2+c_2^2)-2c_1c_2\Big),   \non \\
\Gamma^{\rm ann}(\Omega_c) &=& 6|V_{us}/V_{ud}|^2\,\Gamma_0\,\eta_{\rm 
nspec}\,r_{\Omega_c}(1-x^2)\Big(\tilde B(c_1^2+c_2^2)-2c_1c_2\Big).
\en
The $\Omega_c$ matrix element [see Eq.~(2.18)]
\be
\la\Omega_c|(\bar cs)(\bar sc)|\Omega_c\ra=\,-6|\psi^{\Omega_c}_{cs}(0)|^2
(2m_{\Omega_c})
\en
accounts for the factor 6 in Eq.~(4.9).

   To proceed we employ the Wilson coefficients
\be
c_1(\mu)=\,1.35,~~~~~c_2(\mu)=-0.64
\en
evaluated at the scale $\mu=1.25$ GeV. From Eqs.~(3.21) and (3.22) we obtain
$\tilde B(\mu)\simeq 0.74\tilde  B(\mu_{\rm had})\simeq 0.74$ and $r(\mu)
\simeq 1.36\,r(\mu_{\rm had})$. Repeating the same exercise as the bottom 
baryon case, the results of calculations are exhibited in Table III. We see 
that the lifetime pattern
\be
\tau(\Xi_c^+)>\tau(\Lambda_c^+)>\tau(\Xi_c^0)>\tau(\Omega_c^0)
\en
is in accordance with experiment. It is evident that when nonspectator effects
in semileptonic decay are included, as shown in parentheses in Table III,
the discrepancy between theory and experiment is improved. This lifetime
hierarchy (4.12) is qualitatively understandable. The $\Xi_c^+$ baryon is
longest-lived among charmed baryons because of the smallness of 
$W$-exchange and partial cancellation between constructive and destructive
Pauli interferences, while $\Omega_c$ is shortest-lived due to the
presence of two $s$ quarks in the $\Omega_c$ that renders the contribution of
$\Gamma^{\rm int}_+$ largely enhanced. It is also clear from Table III that, 
although the qualitative feature of the lifetime pattern is comprehensive, the
quantitative estimates of charmed baryon lifetimes and their ratios are 
still rather poor.

\vskip 0.4cm
{{\small Table III. Various contributions to the decay rates (in units of
$10^{-12}$ GeV) of charmed baryons. When nonspectator
effects in semileptonic decay are included, the predictions are shown in
parentheses. Experimental values are taken from \cite{PDG}.}}
{
\begin{center}
\begin{tabular}{|c|c c c c l l l l|} \hline
 & $\Gamma^{\rm dec}$ & $\Gamma^{\rm ann}$ & $\Gamma^{\rm int}_-$ &
$\Gamma^{\rm int}_+$ & ~
$\Gamma_{\rm SL}$ & ~$\Gamma^{\rm tot}$ & ~$\tau(10^{-13}s)$~ & ~
$\tau_{\rm expt}(10^{-13}s)$~ \\  \hline ~
$\Lambda_c^+$~ & ~0.903~ & ~0.858~ & ~$-0.238$~ & & ~0.306~ & ~1.829~ & ~
3.60~ & ~$2.06\pm 0.12$~   \\
$\Xi_c^+$ & 0.903 & 0.042 & $-0.226$ & 0.423 & ~0.306(0.498) & ~1.447(1.639) 
& ~4.55(4.02) & ~$3.5^{+0.7}_{-0.4}$  \\
$\Xi_c^0$ & 0.903 & 0.817 & & 0.423 & ~0.306(0.498) & ~2.448(2.640) & ~
2.69(2.49) & ~$0.98^{+0.23}_{-0.15}$ \\
$\Omega_c^0$ & 0.968 & 0.224 & & 1.256 & ~0.262(0.772) & ~2.710(3.220) & ~
2.43(2.04) & ~$0.64\pm 0.20$  \\
\hline
\end{tabular}
\end{center} }
\vskip 0.4cm

 In order to have a better quantitative description of nonleptonic inclusive
decays of charmed baryons, we shall follow \cite{Altarelli} to assume that 
$\Gamma_{\rm NL}$ scales with $m_{H_c}^5$ instead of $m_c^5$:
\be
&& \Gamma_{\rm NL}(\Lambda_c):\,\Gamma_{\rm NL}(\Xi_c^+):\,\Gamma_{\rm NL}(
\Xi_c^0):\,\Gamma_{\rm NL}(\Omega_c)   \non \\
&& =\Gamma_{\rm NL}^{(0)}(\Lambda_c)\left({m_{\Lambda_c}\over m_c}\right)^5:\,
\Gamma_{\rm NL}^{(0)}(\Xi_c^+)\left({m_{\Xi_c^+}\over m_c}\right)^5:\,
\Gamma_{\rm NL}^{(0)}(\Xi_c^0)\left({m_{\Xi_c^0}\over m_c}\right)^5:\,
\Gamma_{\rm NL}^{(0)}(\Omega_c)\left({m_{\Omega_c}\over m_c}\right)^5,
\en
where $\Gamma_{\rm NL}^{(0)}$ is the nonleptonic decay rate calculated
in the framework of the heavy quark expansion and it has the form
\be
\Gamma_{\rm NL}^{(0)}=\,{G_F^2m_c^5\over 192\pi^3}\Big[a+b/m_c^2+c/m_c^3+{
\cal O}(1/m_c^4)\Big].
\en
To compute the absolute decay width, we introduce a parameter $\lambda$ so 
that
\be
\Gamma_{\rm NL}^{(0)}\longrightarrow \Gamma_{\rm NL}=\,\lambda\Gamma_{\rm NL}
^{(0)}\left({m_{H_c}\over m_c}\right)^5.
\en
Unlike the ansatz (3.23) for bottom hadrons, it will become clear shortly that
$\lambda$ is much less than unity for charmed hadrons.
Applying the prescription (4.15), treating $\lambda,~r,~\tilde B$ as
free parameters and fitting them to the data of charmed baryon lifetimes
\cite{PDG}, we find
\be
\lambda=0.18,~~~~r=1.72,~~~~\tilde B=1.46,
\en
where $r$ and $\tilde B$ are renormalized at $\mu=1.25$ GeV. The numerical
results are summarized in Table IV. Contrary to the previous case, a
prefect agreement with experiment will be achieved if nonspectator effects in
semileptonic decay are not included.

\vskip 0.4cm
{{\small Table IV. Same as Table III except that the ansatz (4.15) has been 
applied to enhance the nonleptonic $c$ quark decay and nonspectator effects.}}
{
\begin{center}
\begin{tabular}{|c|c c c c l l l l|} \hline
 & $\Gamma^{\rm dec}$ & $\Gamma^{\rm ann}$ & $\Gamma^{\rm int}_-$ &
$\Gamma^{\rm int}_+$ & ~
$\Gamma_{\rm SL}$ & ~$\Gamma^{\rm tot}$ & ~$\tau(10^{-13}s)$~ & ~
$\tau_{\rm expt}(10^{-13}s)$~ \\  \hline  ~
$\Lambda_c^+$~ & ~0.960~ & ~2.753~ & ~$-0.884$~ & & ~0.306~ & ~3.136~ & ~
2.10~ & ~$2.06\pm 0.12$~   \\
$\Xi_c^+$ & 1.404 & 0.195 & $-1.231$ & 1.227 & ~0.306(0.837) & ~1.902(2.432) 
& ~3.46(2.70) & ~$3.5^{+0.7}_{-0.4}$  \\
$\Xi_c^0$ & 1.415 & 3.868 & & 1.238 & ~0.306(0.837) & ~6.828(7.358) & ~
0.96(0.89) & ~$0.98^{+0.23}_{-0.15}$ \\
$\Omega_c^0$ & 2.389 & 1.668 & & 5.775 & ~0.262(1.675) & ~10.09(11.51) & ~
0.65(0.57) & ~$0.64\pm 0.20$  \\
\hline
\end{tabular}
\end{center} }
\vskip 0.4cm

Let us examine the fitted parameters (4.16) in more detail. The value $r=1.72$
is fairly reasonable as it implies $|\psi^{\Lambda_c}_{cq}(0)|^2=1.1\times 
10^{-2}{\rm GeV}^3$, which is consistent with those of hyperons and bottom
baryons. Then, does it mean that our previous estimate of $r$ for charmed 
baryons [see
(2.34)] is too small ? In our opinion, the enhancement of $|\psi^{\Lambda_c}
_{cq}(0)|^2$ is likely due to the fact that $|\psi^D_{c\bar q}(0)|^2$ is not
simply equal to ${y\over 12}f_D^2m_D$ with $y=1$ and $f_D\approx 200$ MeV.
We conjecture that a more
realistic value of $y$ is probably close to 3 for charmed baryons and
to unity for bottom baryons. 

   As for the parameter $\tilde B(\mu)$, it is expected to be less than 
unity if the valence-quark approximation is believed to be valid at a 
lower hadronic scale. Therefore, it is not clear to us why $\tilde B(\mu)$ is
larger than unity and what is its implication.
The smallness of
$\lambda$ is attributed to the fact that the inclusive nonleptonic decays of
charmed baryons are not dominated by the $c$ quark decay. Nonspectator effects
of $W$-exchange and Pauli interference terms are expected to be of order
\be
16\pi^2(\Lambda_{\rm QCD}/m_c)^3\sim 0.5-0.7\,,
\en
where the factor $16\pi^2$ is a two-body phase-space enhancement relative to 
the three-body phase space of heavy quark decay.
Realistic calculations (see
Tables III and IV) indicate that nonspectator contributions are comparable to
and even dominate over the $c$ quark decay mechanism. This implies that the
charmed quark is not heavy enough (i.e. the energy release is not sufficiently
large) to make a sensible and meaningful heavy quark expansion. For bottom
hadrons, we see in Sec.~III that at least for the $\Lambda_b$ baryon and $B$
mesons, the nonleptonic decay rate is approximated by
\be
\Gamma_{\rm NL}(H_b)\approx \Gamma^{\rm dec}\left({m_{H_b}\over m_b}\right)^5,
\en
where $\Gamma^{\rm dec}$ is the heavy quark decay rate. However, we find for
charmed baryons that $\Gamma^{\rm dec}(m_{H_c}/m_c)^5$ are 5.36, 7.84, 7.84, 
13.24 (in units of $10^{-12}$GeV), respectively, for $\Lambda_c,~\Xi_c^+,~
\Xi_c^0$ and $\Omega_c$, where $\Gamma^{\rm dec}$ is taken from Table 
III. Therefore, even in the absence of $1/m_c^2$ and $1/m_c^3$ corrections or
even the heavy quark expansion converges, the scaled nonleptonic 
$c$-quark decay rate $\Gamma^{\rm dec}(m_{H_c}/m_c)^5$ already exceeds
the experimental decay widths: 3.20, 1.88, 7.72, 10.28 (in units 
of $10^{-12}$GeV) \cite{PDG}; that is,
\be
\Gamma^{\rm dec}\left({m_{H_c}\over m_c}\right)^5>\Gamma_{\rm tot}(H_c),
\en
except for the $\Xi_c^0$. The presence of large nonspectator contributions 
(see Tables III and IV) will make the discrepancy between 
theory and experiment for decay widths even much worse. Hence, we have to 
introduce a parameter $\lambda\ll 1$ to suppress the absolute rates. However,
since $\lambda$ is an entirely unknown parameter in theory, the recipe of
scaling $\Gamma_{\rm NL}$ with the fifth power of charmed hadron mass is
{\it ad hoc} and does not have the predictive power for the absolute decay 
widths. We conclude that, although the
ansatz (4.13) provides a much better description of lifetime
{\it ratios} for charmed baryons (apart from the annoying parameter 
$\tilde B$),
the prescription (4.15) appears unnatural and unpredictive for describing
the {\it absolute} inclusive decay rates of charmed baryons due to the
presence of the unknown parameter $\lambda$. Since the heavy
quark expansion converges very badly, local duality is thus not testable in 
inclusive nonleptonic charm decay.

\section{DISCUSSIONS AND CONCLUSIONS}
   We have analyzed the lifetimes of bottom and charmed hadrons within the
framework of the heavy quark expansion. Especial attention is paid to the
nonperturbative parameter $\lambda_2^{\rm baryon}$ and four-quark matrix
elements for baryons. We found that the large-$N_c$ relation $\lambda_2^{\rm
meson}\sim N_c\lambda_2^{\rm baryon}$ is satisfactorily obeyed by
bottom hadrons. We have followed \cite{Neubert97} to parametrize the
four-quark matrix elements in a model-independent way. Baryon matrix
elements are evaluated using the NQM and the bag model. The bag-model
estimate for bottom-baryon matrix elements is smaller than that of the NQM
by a factor of $\sim$3. The hadronic parameter $r$ defined in Eq.~(2.27) is
estimated in the NQM to be in the range 0.53 to 0.61 for both bottom and 
charmed baryons. Nonspectator effects in inclusive nonleptonic decays are 
then studied in detail.
The main results of our analysis are as follows.

\begin{itemize}
\begin{enumerate}
\item  Using the charmed quark pole mass fixed from the measured 
semileptonic decay widths of $D^+$ and $D^0$, we have calculated $1/m_Q^2$
nonperturbative corrections to the semileptonic inclusive widths for other
heavy hadrons. We found that while $\Gamma_{\rm SL}(B)$ is very close to 
$\Gamma_{\rm SL}(\Lambda_b)$, $\Gamma_{\rm SL}(D)$ is smaller 
than $\Gamma_{\rm SL}(\Lambda_c)$. The predicted semileptonic decay rates 
for the $B$ meson and the $\Lambda_c$ baryon are in good agreement with 
experiment. This implies that global duality is valid for inclusive
semileptonic decay. For
charmed baryons $\Xi_c$ and $\Omega_c$, there is an additional contribution
to the semileptonic width coming from the constructive Pauli interference
of the $s$ quark. This interference effect is sizeable for the $\Xi_c$
and becomes overwhelming for the $\Omega_c$.

\item The lifetime pattern of the bottom baryons is predicted to be $\tau
(\Omega_b)\simeq\tau(\Xi_b^-)>\tau(\Lambda_b)\simeq \tau(\Xi_b^0)$.
Nonspectator effects due to $W$-exchange and destructive Pauli interference
account for their lifetime differences. The model-independent expression
in the OPE for $\tau(\Lambda_b)/\tau(B_d)$ is given by (3.19), which is 
difficult to accommodate the data without invoking unnaturally too large
values of hadronic parameters. Irrespective of the short $\Lambda_b$ lifetime 
problem, the calculated absolute decay width of the charged $B^-$ meson is 
at least 20\% too small compared to experiment. Since the
predicted $\Gamma_{\rm SL}(B)$ agrees with data, the deficit of the $B$
meson decay rate is blamed on the nonleptonic width.

\item Unlike the semileptonic decays, the heavy quark expansion in inclusive
nonleptonic decay cannot be justified by analytic continuation into the
complex plane and local duality has to be assumed in order to apply the OPE
directly in the physical region. The shorter lifetime of the $\Lambda_b$
relative to that of the $B_d$ meson suggests a significant violation of
quark-hadron local duality. The simple ansatz that $\Gamma_{\rm NL}\to
\Gamma_{\rm NL}(m_{H_b}/m_b)^5$ not only solves the lifetime ratio problem
but also provides the correct absolute decay widths for the $\Lambda_b$
baryon and the $B$ meson. The hierarchy of bottom baryon lifetimes is
modified to $\tau(\Lambda_b)>\tau(\Xi_b^-)>\tau(\Xi_b^0)> \tau(\Omega_b)$:
The longest-lived $\Omega_b$ among bottom baryons in the OPE approach now
becomes shortest-lived. This ansatz can be tested by measuring the $\Xi_b$
lifetime in the near future. More precise measurement of the $B_s$ lifetime
provides another quick and direct test of local duality.

\item The lifetime hierarchy $\tau(\Xi_c^+)>\tau(\Lambda_c)>\tau(\Xi_c^0)
>\tau(\Omega_c)$ is qualitatively understandable in the OPE approach but not
quantitatively. Apart from an annoying feature with the parameter $\tilde B$,
a better description of inclusive decays of charmed baryons
is achieved by scaling $\Gamma_{\rm NL}$ with $m^5_{H_c}$ instead of
$m_c^5$. Contrary to the bottom case, a small parameter $\lambda\ll 1$ has to 
be introduced, namely $\Gamma_{\rm NL}\to \lambda\Gamma_{\rm NL}(m_{H_c}/
m_c)^5$, otherwise absolute decay widths of charmed baryons will be
largely overestimated. Since $\lambda$ is an entirely unknown parameter
in theory, it renders the above prescription unnatural and less predictive.
As the heavy quark expansion in charm decay converges very badly, it is
meaningless to test local duality in nonleptonic inclusive decay of
charmed hadrons.
\end{enumerate}
\end{itemize}

   We conclude that the recipe of allowing the presence of linear $1/m_Q$
corrections by scaling the nonleptonic decay widths with the fifth power
of the hadron mass is operative in the bottom family but becomes unnatural
in charm decay. Can this prescription be justified in a more fundamental way ?
It is interesting to note that a PQCD-based factorization formulism has been
developed for inclusive semileptonic $B$ meson decay \cite{Yu}. This
approach is formulated directly in terms of meson-level kinematics. 
Quark-hadron duality can be tested by comparing results obtained from
quark-level kinematics and those from meson kinematics. The validity of
global duality has 
been demonstrated in the general kinematic region up to ${\cal O}(1/m_Q^2)$;
$1/m_Q$ corrections to inclusive semileptonic widths are indeed
nontrivially canceled out. When this factorization approach is generalized
to nonleptonic decays and to heavy baryons, it is natural to expect that
$\Gamma_{\rm NL}(B)/\Gamma_{\rm NL}(\Lambda_b)\approx (m_B/m_{\Lambda_b})^5$ 
if local duality is violated. Since the application of PQCD and hence the 
factorization scheme of \cite{Yu} to charm decay is very marginal due to
the fact that the charmed hadron scale is not sufficiently large,
the scaling behavior of $\Gamma_{\rm NL}$ with $m^5_{H_Q}$ occurred in the 
bottom decay is no longer anticipated in inclusive nonleptonic decays of
charmed hadrons.

\vskip 0.5 cm
\vskip 0.5cm
\noindent {\bf  Acknowledgments}:~~I am grateful to Hoi-Lai Yu for helpful 
discussions.
This work was supported in part by the National Science Council 
of ROC under Contract No. NSC86-2112-M-001-020.

%
%
\newcommand{\bi}{\bibitem}
%

\end{document}